\documentclass[trackchanges]{aastex631}
\usepackage{CJK}

\shorttitle{AASTeX v6.31 Sample article}
\shortauthors{Hu et al.}

\graphicspath{{./}{figures/}}
\usepackage{soul}
\usepackage{color,xcolor}
\usepackage{amsmath}
\usepackage{boondox-cal}
\usepackage{ulem}

\begin{document}
\begin{CJK*}{UTF8}{gbsn}

\title{Excitation of quasi-periodic fast-propagating waves in the early stage of the solar eruption}

\author[0000-0001-9828-1549]{Jialiang Hu}
\affiliation{Yunnan Observatories, Chinese Academy of Sciences, P.O. Box 110, Kunming, Yunnan 650216, People's Republic of China}
\affiliation{University of Chinese Academy of Sciences, Beijing 100049, People's Republic of China}
\affiliation{Yunnan Key Laboratory of Solar Physics and Space Science, Kunming, Yunnan 650216, People's Republic of China}

\author[0000-0002-5983-104X]{Jing Ye}
\affiliation{Yunnan Observatories, Chinese Academy of Sciences, P.O. Box 110, Kunming, Yunnan 650216, People's Republic of China}
\affiliation{Yunnan Key Laboratory of Solar Physics and Space Science, Kunming, Yunnan 650216, People's Republic of China}
\affiliation{Yunnan Province China-Malaysia HF-VHF Advanced Radio Astronomy Technology International Joint Laboratory, Kunming, Yunnan 650216, People’s Republic
of China}

\correspondingauthor{Jing Ye, Zhixing Mei }
\email{yj@ynao.ac.cn, meizhixing@ynao.ac.cn}

\author[0000-0002-8077-094X]{Yuhao Chen}
\affiliation{Yunnan Observatories, Chinese Academy of Sciences, P.O. Box 110, Kunming, Yunnan 650216, People's Republic of China}
\affiliation{University of Chinese Academy of Sciences, Beijing 100049, People's Republic of China}
\affiliation{Yunnan Key Laboratory of Solar Physics and Space Science, Kunming, Yunnan 650216, People's Republic of China}

\author[0000-0001-9650-1536]{Zhixing Mei}
\affiliation{Yunnan Observatories, Chinese Academy of Sciences, P.O. Box 110, Kunming, Yunnan 650216, People's Republic of China}
\affiliation{Yunnan Key Laboratory of Solar Physics and Space Science, Kunming, Yunnan 650216, People's Republic of China}

\author[0009-0002-6808-5330]{Zehao Tang}
\affiliation{Yunnan Observatories, Chinese Academy of Sciences, P.O. Box 110, Kunming, Yunnan 650216, People's Republic of China}
\affiliation{University of Chinese Academy of Sciences, Beijing 100049, People's Republic of China}
\affiliation{Yunnan Key Laboratory of Solar Physics and Space Science, Kunming, Yunnan 650216, People's Republic of China}

\author[0000-0002-3326-5860]{Jun Lin}
\affiliation{Yunnan Observatories, Chinese Academy of Sciences, P.O. Box 110, Kunming, Yunnan 650216, People's Republic of China}
\affiliation{University of Chinese Academy of Sciences, Beijing 100049, People's Republic of China}
\affiliation{Yunnan Key Laboratory of Solar Physics and Space Science, Kunming, Yunnan 650216, People's Republic of China}
\affiliation{Center for Astronomical Mega-Science, Chinese Academy of Sciences, Beijing 100012, People's Republic of China}

\begin{abstract}

We propose a mechanism for the excitation of large-scale quasi-periodic fast-propagating magnetoacoustic (QFP) waves observed on both sides of the coronal mass ejection (CME). Through a series of numerical experiments, we successfully simulated the quasi-static evolution of the equilibrium locations of the magnetic flux rope in response to the change of the background magnetic field, as well as the consequent loss of the equilibrium that eventually gives rise to the eruption. During the eruption, we identified QFP waves propagating radially outwards the flux rope, and tracing their origin reveals that they result from the disturbance within the flux rope. Acting as an imperfect waveguide, the flux rope allows the internal disturbance to escape to the outside successively via its surface, invoking the observed QFP waves. Furthermore, we synthesized the images of QFP waves on the basis of the data given by our simulations, and found the consistence with observations. This indicates that the leakage of the disturbance outside the flux rope could be a reasonable mechanism of QFP waves.

\end{abstract}



\section{Introduction} \label{1}

Solar atmosphere is a highly structured medium capable of supporting various types of waves, which might belong to the families of the Magnetohydrodynamic (MHD) waves in the solar corona including fast-mode \citep{2001MNRAS.326..428W,2002MNRAS.336..747W,2003A&A...412L...7N} and slow-mode waves \citep{2000A&A...355L..23D,2009ApJ...700.1716W,2002ApJ...580L..85O} , and they have been extensively investigated by many authors (e.g., see also  \citealt{2020SSRv..216..136L, 2020ARA&A..58..441N, 2021SSRv..217...76B, 2021SSRv..217...73N, 2021SSRv..217...34W}). They impose additional constraints on the explanation or understanding of fundamental physical processes occurring in solar activity, such as particle acceleration and magnetic energy release \citep{2018SSRv..214...45M}. In addition, these perturbation may interact with the plasma and the magnetic filed nearby, delivering the energy to the background atmosphere, which may eventually compensate the radiative losses of thermal energy in the atmosphere \citep{1994ApJ...435..502P, 2020SSRv..216..140V}.

Global large scale EUV waves, usually observed in the EUV wavelengths, are considered a ubiquitous phenomenon \citep{1998GeoRL..25.2465T,2009ApJ...700.1716W,2010cosp...38.1800L} associated with the disturbance to the atmosphere by the eruption. They typically manifest as bright, ring-like structures that gradually propagate outward, sweeping across a significant portion of the solar surface and leaving behind a dimming area \citep{2009ApJS..183..225T,2012SCPMA..55.1316M}. In their propagation, refraction and reflection may occur to them as they go through regions of different densities or  through the boundaries separating magnetic fields of different topologies, like that between active regions and coronal holes \citep{2009ApJ...691L.123G,2019ApJ...873...22S,2009ApJ...700.1716W,2015ApJ...805..114W,2019MNRAS.490.2918X}. Furthermore, if the wave is energetic enough, the nearby magnetic fields of opposite polarity could even be pushed to approach each other, leading to the occurrence of magnetic reconnection \citep{2020ApJ...905..150Z}.

The quasi-periodic wave train (QFP, \citealt{2011ApJ...736L..13L}) is  an intriguing phenomenon of the disturbance to the corona that was observed for the first time by the Atmospheric Imaging Assembly (AIA) onboard the Solar Dynamics Observatory (SDO, \citealt{2012SoPh..275...17L}). It has been interpreted as the fast-mode magnetoacoustic wave \citep{2003A&A...409..325C}, characterized by continuous, narrow arc-like structures that propagate rapidly at speed up to 1416~km~s$^{-1}$\citep{2013SoPh..288..585S,2017ApJ...844..149K}.  A QFP is most apparent in AIA 171~\AA~ (the associated emission produced by Fe {\sc ix} ions and peaking at temperature of around 0.8~MK), and is often closely related to solar flares. Typically, QFPs first appear more than 100 Mm away from the flare center, possibly due to the time it takes for the disturbance to grow and become detectable after propagating a certain distance \citep{1983Natur.305..688R}. Some QFP waves exhibit periods consistent with that of the oscillating component of the flare light curve, which might imply that the flare invokes the QFP waves \citep{2019ApJ...871L...2M}. The mechanisms behind the formation of quasi-periodic pulsations in flares include the dispersion of perturbations, magnetic configuration resonance, nonlinear processes during magnetic reconnection, and oscillations leaking from the other atmospheric layers into the solar corona \citep{2009SSRv..149..119N}.

The magnetic configuration resonance is typically associated with long-period global oscillations of coronal loops and does not excite QFPs, while the other three mechanisms can excite QFPs. However, some observations suggest that QFPs with multiple narrow arc-like wavefronts are also related to the coronal mass ejection (CME) and appear ahead of CME, indicating that CME is also a possible trigger of the QFP wave \citep{2021arXiv210902847Z}. \cite{2012ApJ...753...52L} discovered QFPs on the inner side of a global-scale EUV wave, and noticed that QFPs propagated on outside of CMEs, propagating a distance of over 0.5 solar radius on the solar surface, at velocities  decreasing from 1400 to 650 ~km~s$^{-1}$. They found that QFPs originated from the same location, but more significant in the lower corona at heights less than 110 Mm, and their brightness rapidly attenuated as the altitude increases. Due to the limitation to present observational techniques, the origin of QFPs remains unclear. Several mechanisms have been proposed to explain the relationship between QFPs and  CMEs by \cite{2021ApJ...911L...8W} and \cite{2019ApJ...873...22S} or flares by \cite{1984ApJ...279..857R},  \cite{2004MNRAS.349..705N}, \cite{2014A&A...569A..12N}, \cite{ 2011ApJ...740L..33O}, \cite{2018ApJ...860...54O}, but still need to be further investigated.


Many authors have investigated QFP waves in the corona via numerical simulations. \cite{2015ApJ...800..111Y} found that the QFP wave could result from collisions of the plasmoids produced in the current sheet by the tearing mode instability with the magnetic field nearby, and propagated  at speed up to 1000 ~km~s$^{-1}$.  \cite{2016ApJ...823..150T} revealed that QFPs are excited by the local oscillation on the top of flare loops, which is governed by magnetic reconnection that takes place in the periodical fashion. \cite{2021ApJ...909...45Y}  discovered QFPs at the trailing edge of CME and suggested a potential connection to the turbulence.

On the other hand, for MHD wave, magnetic field structures filled with plasma may act as effective waveguides \citep{1983Natur.305..688R}. Based on the difference in the speeds of the waves inside and outside the waveguide, the structure of the waveguide as well as the internal wave could be classified into two categories. The first one corresponds to the case in which the speed of the internal wave is lower than that of the fast magnetoacoustic wave outside the waveguide, so the internal wave is well confined in the waveguide. The second one is of the case that the internal wave propagates faster than the fast magnetoacoustic wave outside the waveguide, so the internal wave could leak to outside of the waveguide, and propagates outward continuously. \cite{2018ApJ...860...54O} performed simulations of QFPs using a three-dimensional MHD model and successfully obtained various parameters for QFPs, including waveform, amplitude, wavelength, and propagating speed, as well as duplicated the reflection of QFP on the closed magnetic field structures. They also found that the speed and energy flux density of QFPs rapidly decrease with the altitude, which could be ascribed to the divergent characteristics of the funnel-shaped magnetic structure and rapid weakening of the magnetic field at higher altitude. Furthermore, \cite{2022SoPh..297...20S} noted an interesting feature of QFPs, namely, wave amplitude grows initially, and decreases later on during propagation. This phenomenon may result from the combination of the growth in the amplitude of the wave and the expanding feature of the coronal magnetic field configuration that serves as waveguides. Here the growth of the wave amplitude is believed to result from the gravity stratification of the atmosphere .

Regarding the leakage of the wave from the waveguide-like magnetic structure in the corona, the seminal analytical work on leaky waves was conducted by \cite{1986SoPh..103..277C}, and a comparative study of the leaky and trapped regimes was performed by \cite{2012ApJ...761..134N}. \cite{2013A&A...560A..97P,2014A&A...568A..20P}  studied the leakage of the wave from the funnel-shaped magnetic structure. They saw the leaking wave propagating outward, exhibiting multiple ``wing"-shaped wavefronts, which might account for the QFP wave observed in the eruption. \cite{2023ApJ...943L..19S} performed three-dimensional MHD simulations to study the impact of the kink-like velocity perturbation to the flare loop (modeled by the straight cylinder). Their results revealed that the straight cylinder could behave as waveguides, allowing long-period (57~s period) and short-period (5.8~s period) oscillations to occur within the waveguide. They used a fast gyrosynchrotron code to produce synthetic signals in the microwave band, explaining the generation of quasi-periodic pulsations (QPPs) in solar flares. It is worth noting that although not explicitly mentioned in their study, their results suggest that when short-period oscillation signals pass through the boundary of the cylinder, a portion of the oscillation leaks outside the cylinder and forms multiple wavefronts, which is similar to the multiple QFP waves observed in the solar eruption. Analyzing the observational data from the 1.6-meter Goode Solar Telescope, \cite{2023NatAs...7..856Y} discovered persistent transverse waves in the chromospheric umbral fibrils of a sunspot with strong magnetic field. They reproduced these waves using a two-fluid MHD simulation. Interestingly, their results indicate that perturbations confined within the chromospheric fibrils (acting as waveguide structures) oscillated back and forth, and leaked out at the boundaries, giving rise to multiple wavefront signals.

In order to investigate the QFP phenomena and mechanisms for their generation during solar eruptions, we will first examine the evolution in the coronal magnetic structure including a magnetic flux ropes through a set of quasi-static states to eventually losing the equilibrium. Then, we follow the consequent dynamic evolution in the system, investigate the disturbance to the background environment and the occurrence of QFP waves in this process, look into the physical property and mechanism triggering the QFP wave. We will provide a detailed introduction of the model and the relevant numerical techniques used in the numerical simulations in Section \ref{part2}, and display our results in Section \ref{part3}. Synthetic images deduced according to the numerical experiments and the associated comparisons with observations will be given in Section \ref{part4}. Finally, we summarize this work in Section \ref{part5}.

\section{Descriptions of the numerical method}\label{part2}

We utilize the MPI-AMRVAC code \citep{2012JCoPh.231..718K,2014ApJS..214....4P,2018ApJS..234...30X}  to study the evolution in the coronal magnetic configuration through a set of quasi-static equilibrium states to the loss of equilibrium. Our calculations are involved in solving the following MHD equations, which include the gravity and the thermal conduction:

\begin{equation}
  \partial_{t}\rho+\bigtriangledown \cdot (\rho \textbf{v})=0\label{continum},
\end{equation}

\begin{equation}
 \partial_{t}(e)+\nabla  \cdot [(e+p+\frac{1}{2\mu_{0}}|\textbf{B}|^{2})-\frac{1}{\mu_{0}}(\textbf{v}\cdot \textbf{B})\textbf{B}]
  = \rho \textbf{g} \cdot \textbf{v} + \nabla  \cdot [\kappa_{\parallel}(\nabla  T\cdot \hat{\textbf{B}})\hat{\textbf{B}}],
\end{equation}

\begin{equation}
  \partial_{t}(\rho \textbf{v})+\bigtriangledown \cdot [\rho\textbf{v}\textbf{v}+(p+\frac{1}{2}|\textbf{B}|^{2})\textbf{I}-\textbf{B}\textbf{B}] =\rho \textbf{g} ,
\end{equation}

\begin{equation}
 \partial_{t}\textbf{B}  =   \bigtriangledown \times (\textbf{v} \times \textbf{B}-\eta \bigtriangledown \times \textbf{B}),
\end{equation}
where $\rho$, $\textbf{v}$, $\textbf{B}$, $p$, $e$, and $T$ are density, velocity, magnetic field, pressure, energy, and temperature, respectively. In the description of thermal conduction in the plasma, we employ the classical Spitzer model \citep{1962pfig.book.....S} with $\kappa_{\parallel}$= 9000 W~m$^{-1}~$K$^{-1}$ for the coefficient of the thermal conductivity parallel to the magnetic field. The gravity is $\textbf{g}(y)=g_{0}/(1+y/r_{0})^{2} \hat{\textbf{y}}$ with $g_{0}$ ($g_{0}=274$ m~s$^{-2}$) being the value of the gravity on the solar surface, $r_{0}$ ($r_{0}=6.9551\times10^{8}$~m) the radius of the Sun, and $\hat{\bold{y}}$ the unit vector in the $y$-direction. Due to numerical diffusion in calculations, we set $\eta$ to 0 for simplicity. We note here that the thermal conduction parallel to the magnetic field is included in our calculations. The point of doing so is that the thermal conduction in the coronal environment is a dominant process of the energy transportation, and that the field-aligned conduction is powerful compared to that perpendicular to the field in the corona. Therefore including a field-aligned thermal conduction in calculations here is of consideration for the model to be as realistic as possible.

In addition, the energy $e$ and the pressure $p$ are given as:
\begin{equation}
  e = \frac{1}{2} \rho v^{2} + \frac{p}{\gamma-1}+\frac{B^{2}}{2\mu_{0}}\label{energy},
\end{equation}
\begin{equation}
 p = 2 k_{b}\rho(y) T(y)/m_{H},
\end{equation}
where $\gamma=5/3$. To suppress spurious oscillations in our computations, we employ a third-order asymmetric TVD limiter and utilize a  ``three-step" scheme of the third-order predictor-corrector. Regarding the divergence-free of magnetic field, we apply the ``typedivbfix=linde'' condition in practical calculations to maintaining the numerical conservation of the magnetic field divergence ($\nabla \cdot \textbf{B}=0$). According to Equation (\ref{energy}), in low-beta plasma atmospheres, numerical errors in the magnetic energy may potentially exceed the internal energy, leading to negative values for the internal energy or gas pressure when solving the complete energy equation. To address this issue, we simultaneously solve both the complete energy equation and the internal energy equation.

\subsection{the gravitationally stratified atmosphere}

Calculations are conducted in the Cartesian coordinate system $(x, y)$, with the origin located at the bottom of the photosphere. The $x$-axis lies on bottom of the photosphere, and the $y$-axis is the radial direction away from the Sun. The simulation domain spans over $-1.6L_{0} \leq x \leq 1.6L_{0}$ and $0 \leq y \leq 3.2L_{0}$, where $L_{0} = 2\times10^{10}$~cm. In our simulation, we employ a two-layer atmospheric structure to model the solar atmosphere, where the region with higher density ($y < h_p$) represents the photosphere, and the region with lower density ($y > h_p$) represents the corona. The temperature of the gravitationally stratified atmosphere is set to:

\begin{equation}
T(y)=\left\{
\begin{aligned}
 &T_{p} = 2\times10^{3} ~\rm K,  \quad & 0 \leq y \leq h_{p} &,  \\
 &T_{c} = 10^{6} ~\rm K ,  \quad & h_{p} < y &.&
\end{aligned}
\right.
\end{equation}
For the photospheric layer, we follow the setting of \cite{2012SCPMA..55.1316M}, and take $h_p=2.5\times 10^7$~cm. The isothermal property of the atmosphere in the low corona yields the gas pressure:

\begin{equation}
p(y)=\left\{
\begin{aligned}
 &p_{p}\exp\left[-\frac{y}{\lambda_{p}(1+y/r_{0})}\right],  \quad & 0 \leq y \leq h_{p} &,   \\
 &p_{c}\exp\left[-\frac{y-h_{p}}{\lambda_{c}(1+y/r_{0})(1+h_{p}/r_{0})}\right],  \quad & h_{p} < y &,
\end{aligned}
\right.
\end{equation}
where $p_{c}$ and $p_{p}$ are the pressure at the bottom of the corona and the bottom of the photosphere, respectively, with $p_{p}=p_{c}\exp\left[h_{p}/\lambda_{p}/(1+h_{p}/r_{0})\right]$. Here, $\lambda_{p}$ and $\lambda_{c}$ are the pressure scale heights for the photosphere and the corona, respectively:

\begin{equation}
 \lambda_{p} = \frac{2+3H_{ea}}{1+4H_{ea}}\frac{k_{b}T_{p}}{m_{p}g_{0}},
\end{equation}
\begin{equation}
 \lambda_{c}  = \frac{2+3H_{ea}}{1+4H_{ea}}\frac{k_{b}T_{c}}{m_{p}g_{0}},
\end{equation}
where $H_{ea}$, $k_{b}$, and $m_{p}$ represent the helium abundance, Boltzmann constant, and proton mass, respectively. The density distribution could be determined according to the isothermal property of the atmosphere.

\subsection{The structure of magnetic field and filament}

Following \cite{1993ApJ...417..368I}, we model the filament with a current-carrying magnetic flux rope levitating in the corona. To ensure a smooth transition of the parameters from the flux rope to the background corona, we impose a thin shell outside the flux rope with a radius of $R$. This shell is known as the PCTR (prominence-corona transition region, e.g., see also \citealt{2012MNRAS.425.2824M}), with a thickness of  $\Delta = R/5$. Based on the distance $r_{-}$ from any point in space to the center of the flux rope ($r_{-} = \sqrt{x^{2}+(y - h)^{2}}$, where $h$ is the initial height of the flux rope), the entire simulation domain is divided into three regions: the interior of the flux rope (Ar1), for $0\leq r_{-}< (R-\Delta/2)$; the PCTR (Ar2), for $(R-\Delta/2) \leq r_{-} \leq (R+\Delta/2)$; and the exterior of the flux rope (Ar3), for $r_{-}>(R-\Delta/2)$. According to \cite{2012MNRAS.425.2824M}, the current density of the flux rope is:

\begin{equation}
j_{z}(r_{-})=\left\{
	\begin{aligned}
	&j_{0},\quad & r_{-} \in Ar1, \\
	& \frac{j_{0}}{2}\left\{\cos\left[\frac{\pi(r_{-}-R+\Delta/2)}{\Delta}\right] + 1\right\}& r_{-} \in Ar2,\\
	&0,\quad &r_{-} \in Ar3,
	\end{aligned}
	\right.
\end{equation}
where $j_{0}$ is a constant.

The initial magnetic field results from three sources: the internal current of the flux rope centered at $y=h$, the mirror current centered at $y=-h$, and the buried magnetic quadrupoles beneath the photosphere at $y=-d$. So the initial magnetic field is described by:

\begin{equation}
B_{x} = B_{\phi}(r_{-})(y-h)/r_{-} - B_{\phi}(r_{+})(y+h)/r_{+} -B_{\phi}(R+\Delta/2)Md(R+\Delta/2)(y+d)(3x^{2}-(y+d)^2)/r_{d}^{6},
\end{equation}

\begin{equation}
B_{y} = -B_{\phi}(r_{-})x/r_{-} + B_{\phi}(r_{+})x/r_{+} -B_{\phi}(R+\Delta/2)Md(R+\Delta/2)x(-x^{2}+3(y+d)^2)/r_{d}^{6},
\end{equation}
where
\begin{equation}
B_{\phi}(r)=\left\{
	\begin{aligned}
	&-j_{0}J r/2,& \quad  r\in Ar1 \\
	&-j_{0}J((R-\Delta/2)^{2}/2-(\Delta/\pi)^{2}+r^{2}/2+
	(\Delta R/\pi)sin[\pi(r-R+\Delta/2)/\Delta]+\\&(\Delta/\pi)^{2}cos[\pi(r-R+\Delta/2)/\Delta])/(2r),& r \in Ar2\\
	&-j_{0}J[R^{2}+(\Delta/2)^{2}-2(\Delta/\pi)^{2}]/(2r),& \quad r \in Ar3
	\end{aligned}
	\right.
\end{equation}

\begin{equation}
   r_{+}=\sqrt{x^{2}+(y + h)^{2}}, 	
\end{equation}
\begin{equation}
   r_{d} =\sqrt{x^{2}+(y +d)^{2}},
\end{equation}
\begin{equation}
   M = 125\sigma /32,
\end{equation}
with $r_{+}$ and $r_{d}$ being the distances from any point in space $(x, y)$ to the mirror current and the quadrupole, respectively, and $\sigma$ denoting the relative strength of the background magnetic field (interested readers refer to \citealt{1990JGR....9511919F}).

In the magnetohydrostatic equilibrium, the initial gas pressure within the flux rope is given by:

\begin{equation}
p_{f}(y)= p(y) - \int_{r_{-}}^{\infty } j_{z}(r^{\prime})B_{\phi}(r^{\prime})dr^{\prime}+p_{Bz}.\label{pressure}
\end{equation}
The second term on the right-hand side of equation (\ref{pressure}) is the pressure generated by the azimuthal magnetic field, and the third term on the right-hand side is the pressure generated by the axial magnetic field. The external equilibrium condition of the flux rope determines the intensity of the electric current within the flux rope, while the pressure due to the azimuthal magnetic field inside the flux rope governs its internal equilibrium. Prior to numerical experiments, we adjust the axial magnetic pressure according to specific requirements to regulate the internal pressure of the flux rope so that its temperature and density could be as close as possible to those in the true corona.

To smoothly connect the temperature inside the flux rope to that in the background, we set the temperature distribution as follows:

\begin{equation}
T(r_{-})=\left\{
	\begin{aligned}
	&T_{f}, \quad  & r_{-} \in Ar1 \\
	&(T_{c}-T_{f})(r_{-}-R+\Delta/2)/\Delta+T_{f}, \quad  &r_{-} \in Ar2
	\end{aligned}
	\right.
\end{equation}
where $T_f=5\times10^4$ K and $T_c=10^6$ K, the density distribution inside the flux rope can be obtained through the ideal gas law.

\begin{figure}[t]
\centerline{\includegraphics[width=1\textwidth,clip=]{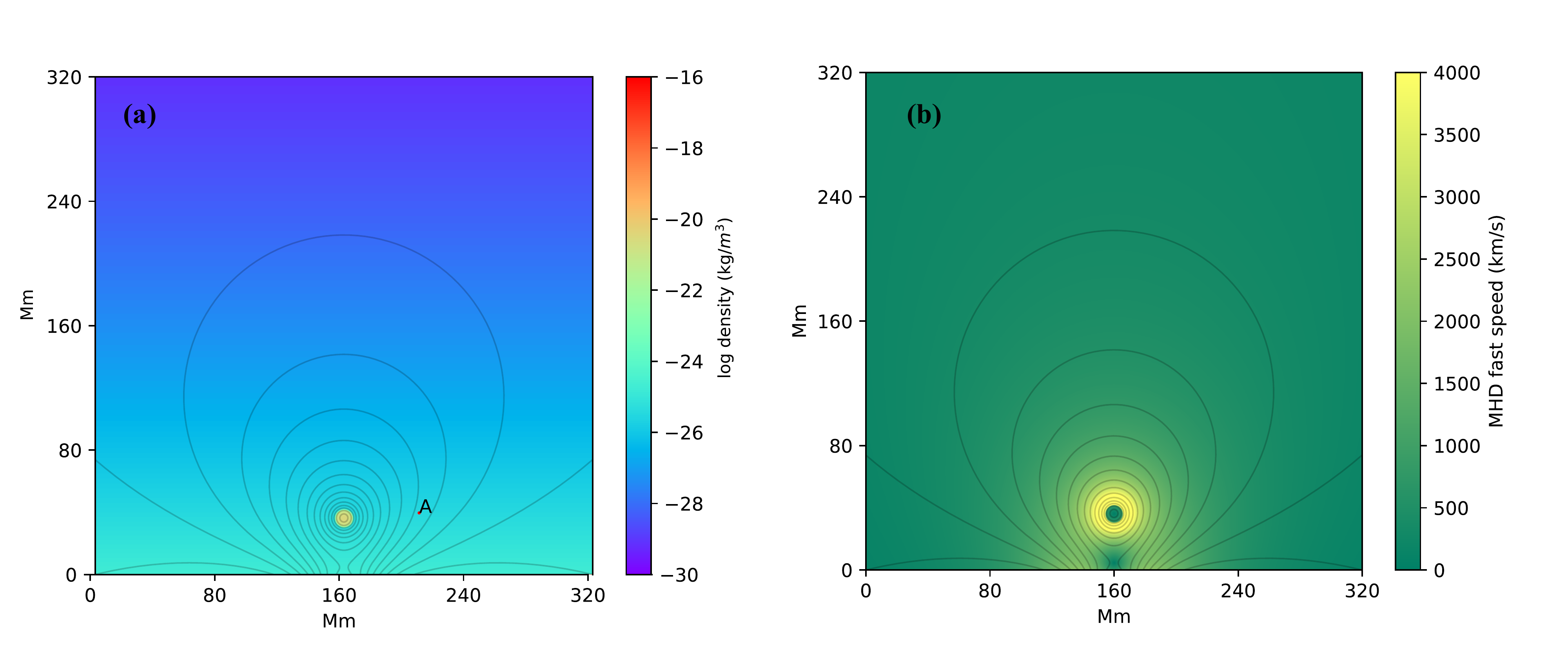}}
\caption{Initial distributions of (a) the plasma density and (b) fast speed. The black solid curves are the magnetic field lines and point A in (a) is selected for analyzing the period of the QFP as shown later in Fig \ref{fig:period}. }
\label{fig:initial}
\end{figure}

Since we need to study the gradual evolution in the background magnetic field driving the coronal magnetic field to evolve through a set of stable equilibrium to a critical state in a quasi-static fashion, to ensure consistency across all sets of numerical experiments, the simulation grid for each experiment is uniformly set to $3840\times3840$. Figure \ref{fig:initial}(a) illustrates the initial magnetic configuration and the density distribution in the gravitationally stratified atmosphere, Figure \ref{fig:initial}(b) displays the distribution of fast speed. We notice that the speed around the flux rope is almost uniform, which allows the wave around the flux rope to propagate almost isotropically. This issue will be further discussed later.

\section{Results} \label{part3}

\begin{figure}[t]
\centerline{\includegraphics[width=1\textwidth,clip=]{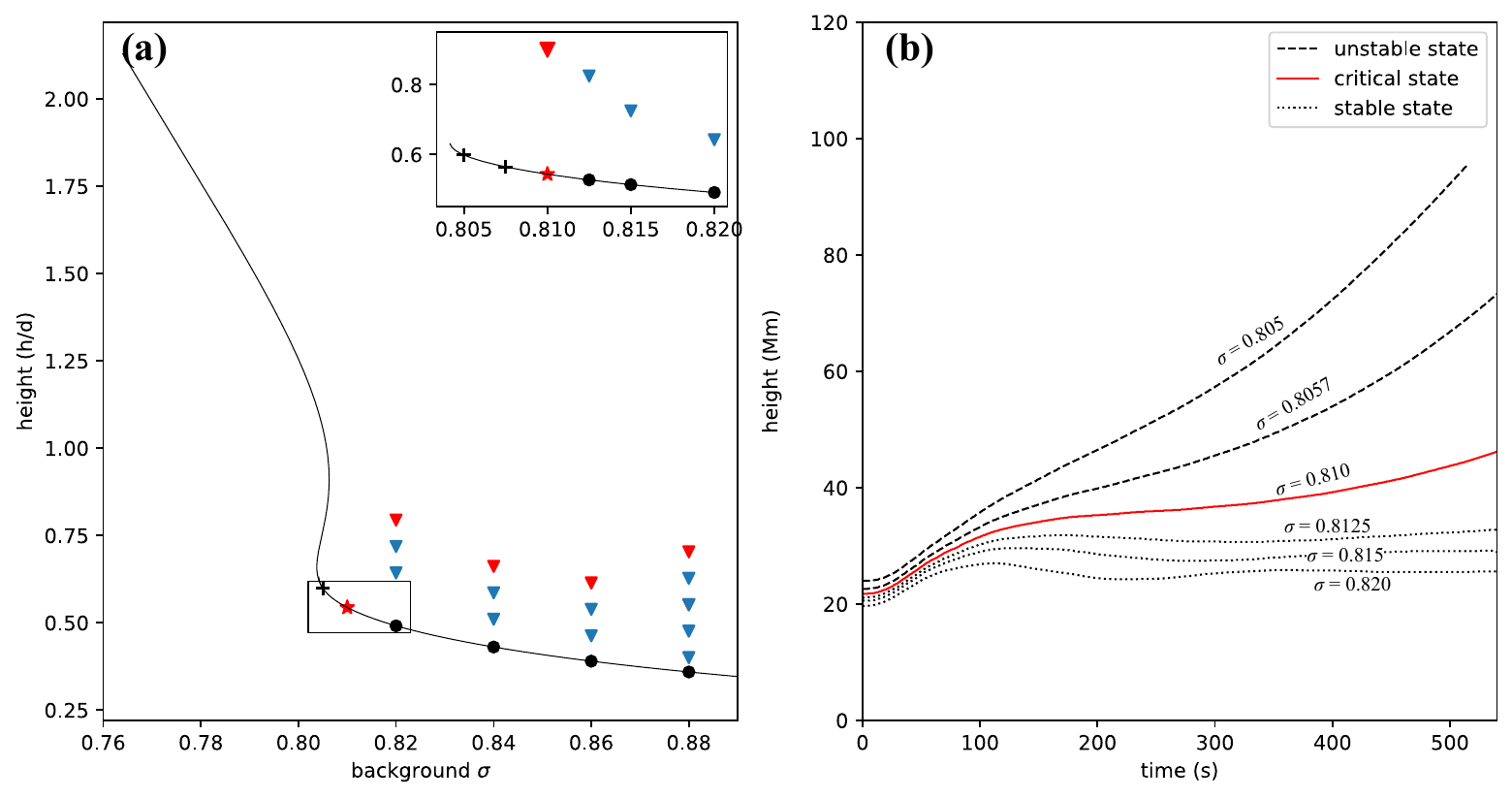}}
\caption{Variations of the equilibrium height of the flux rope versus the strength of the background field. The solid line in panel (a) is the equilibrium curve that was duplicated from the analytical solution of  \cite{1993ApJ...417..368I},  and the black cross marks the critical point. All the solid dots on the equilibrium curve are used as the initial location of the flux rope and the corresponding strength of the background field for numerical experiments. Blue triangles mark positions where the flux rope finds the equilibrium in numerical experiments, while red triangles outline the boundary above which no equilibrium location exists for the flux rope in numerical experiments. The inset in (a) displays more details of equilibrium locations obtained in both theory (continuous curve) and numerical experiments (blue triangles). The flux rope at locations marked by crosses on the curve are still in equilibrium in theory, but not in numerical experiments. Instead the red asterisk and the red triangle of the same value of $\sigma$ mark the critical point for the numerical experiment. Panel (b) shows the temporal evolution of the flux rope location in the corresponding experiments with the initial locations marked by solid dots, asterisk, and crosses on the theoretical curve displayed in the inset of (a). }
\label{fig:catas}
\end{figure}

\cite{1993ApJ...417..368I} studied analytically the change in the background magnetic field driving the coronal magnetic field structure to evolve through a set of stable equilibrium states to a critical state in a quasi-static approach. They described the equilibrium height $h$ of the flux rope as a function of the background magnetic field $\sigma$ (see the black solid line in Figure \ref{fig:catas}(a)). Before investigating the quasi-periodic fast-propagating (QFP) wave in the early stage of the solar eruption, it is essential to verify whether the process investigated by \cite{1993ApJ...417..368I} can indeed occur. In other words, we need to confirm that the coronal magnetic configuration would indeed lose the equilibrium after evolving quasi-statically through a set of equilibrium states to the critical state in response to the change in the background field. This is important for two reasons: first, an unstable state should evolve from the stable state, so the magnetic configuration in reality needs to be in equilibrium for a while prior to the eruption. Second, numerical experiments starts with a configuration in non-equilibrium might bring unphysical issues into calculations, and yielding the unrealistic results and conclusions.

Figure \ref{fig:catas}(a) shows the continuous curve taken from \cite{1993ApJ...417..368I}, depicting the equilibrium height $h$ of the magnetic flux rope as a function of the strength of the background field $\sigma$. Due to discrepancies between analytical and numerical solutions, when we tentatively take the parameters from the continuous curve (black dots) for the initial magnetic configuration in our numerical experiments, it is not surprising that the equilibrium locations (as indicated by the row of blue inverted triangles located at the bottom) of the flux rope in the numerical experiment are not located on the continuous curve, instead the equilibrium locations determined in the numerical experiment always deviate from those deduced from the analytical solution. Moreover, the closer the tentative positions are to the critical point on the analytic curve (near the black cross), the larger the deviation is. The four black dots on the curve correspond to $\sigma=0.88,0.86,0.84$, and 0.82 respectively. The results indicate that when the magnetic flux rope is initially located at heights corresponding to these four points on the equilibrium curve, it undergoes oscillations for a while and eventually finds a stable equilibrium position at the height marked by the triangles closest to the black equilibrium curve after self-adjustment. This demonstrates that the stable equilibrium states and the associated quasi-static evolution in the configuration described by \cite{1993ApJ...417..368I} can indeed be realized. Driven by the change in the photospheric magnetic field and plasma, the system evolves quasi-statically from one stable equilibrium state to another.

 However, a change occurs near the critical point on the equilibrium curve (marked by the cross). The critical point is the joint where the stable section on the equilibrium curve connects to the unstable one. In reality and numerical experiment, no magnetic configuration could exist if its parameters are located on the unstable section of the equilibrium curve. So the loss of equilibrium in the magnetic configuration inevitably occurs as the system evolves to the critical point along the stable section of the equilibrium. In this study, we found that when the initial position is located at point $\sigma=0.80$ (the location marked by symbol ``+'' in Figure \ref{fig:catas}(a)), the flux rope could no longer find equilibrium location through self-adjustment during the consequent evolution, and the motion of the flux rope becomes highly dynamic.

Evolution behaviors of the system at $\sigma=0.82$ and $\sigma=0.80$ suggest the existence of a critical value between these two $\sigma$ values, to which the corresponding magnetic configuration reaches to the critical state where stable and unstable equilibria join. To locate the critical point, we conducted four sets of test experiments between $\sigma=0.82$ and $\sigma=0.80$, using $\sigma$= 0.815, 0.8125, 0.810, and 0.8075. The change in the height of the flux rope versus time for each group of experiments is shown in Figure \ref{fig:catas}(b). From the figure, we can see that when $\sigma$ = 0.820, 0.815, and 0.8125, the flux rope quickly finds a new equilibrium several rounds of oscillation. When $\sigma$ = 0.8075 and 0.805, the flux rope fails to find an equilibrium and rapidly moves upward. When $\sigma=0.810$, the flux rope oscillates and remains at a certain height for a while but eventually loses its equilibrium. This indicates that the catastrophe occurs at this point and it is the critical point found in our numerical experiments, denoted by the red pentagram on the analytical curve in Figure \ref{fig:catas}(a). The height at which the flux rope could stay for a while in the numerical experiment is marked by the red triangle in the inset of Figure \ref{fig:catas}(a). Thus, connecting those triangles closest to the equilibrium curve in Figure \ref{fig:catas}(a) yields the equilibrium curve for our numerical experiments. Figure \ref{fig:catas}(a) also shows that the equilibrium curve obtained from numerical experiments in the present work differs slightly from the analytical results given by \cite{1993ApJ...417..368I}. This difference is primarily due to the existence of the gas pressure, gravity, and numerical dissipation in simulations (e.g., see also \citealt{2022ApJ...933..148C}). However, as expected, these issues do not introduce significant deviations between the numerical and the analytical results because the lower corona is very close to a force-free environment.

Moreover,  in order to find the regions near the equilibrium curve where the flux rope can stably stay, we performed a set of test calculations, and roughly outlined a range of values of the relevant parameters that govern the equilibrium property of the system. We find that the system could eventually reach the equilibrium if the initial value of the relevant parameters were taken in the region below the red triangles and above the solid curve shown in Figure \ref{fig:catas}(a). If the initial values were taken somewhere below the solid curve, the flux rope will move upward and find the equilibrium position in that region as well. The system will never find the equilibrium as the initial value is taken at or above the red triangle, or in the region at the left to the red asterisk on the curve.

So far, we have confirmed that the magnetic configuration shown in Figure \ref{fig:initial} (a) can evolve, in response to the change in the background field, quasi-statically through a set of stable equilibrium configurations to the critical configuration, and then loses its equilibrium, triggering the eruption. The main purpose of this work is to investigate the wave phenomenon in the eruption, so to save computational resources, we directly start our numerical experiments at $\sigma=0.81$ and focus on the issues related to the waves occurring in the eruption.

\subsection{The global evolution} \label{evo}

\begin{figure}[t]
\centerline{\includegraphics[width=1\textwidth,clip=]{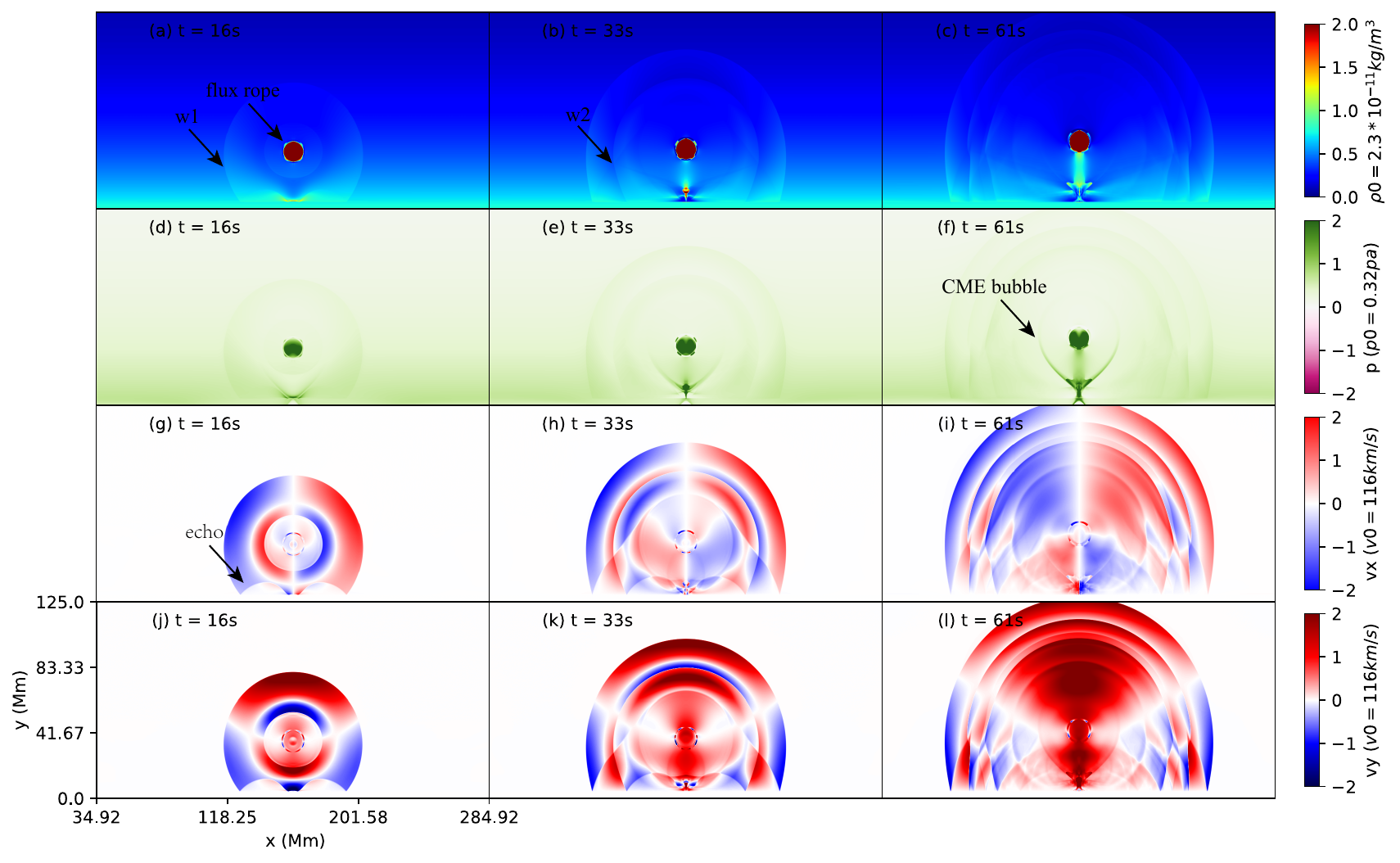}}
\caption{Evolutionary features of quasi-periodic fast propagating (QFP) waves at different times. The rows represent density, pressure, horizontal velocity, and vertical velocity, respectively. The arrows indicate the features described in section \ref{evo}, where ``w1" and ``w2" refer to the first and second wavefronts, respectively.}
\label{fig:evolution}
\end{figure}

Figure \ref{fig:evolution} illustrates evolutionary features in the system at different times. In the early stages of evolution at $t = 16$~s, we see an arc-shaped wavefront (labeled as ``w$_{1}$") that includes the flux rope propagating outward while expanding and extending in various directions simultaneously. As it extends to the bottom boundary, an echo occurs as indicated by black arrows in Figure \ref{fig:evolution}(g), a phenomenon commonly reported in both numerical experiments and observation (e.g., see also \citealt{2019MNRAS.490.2918X,2009ApJ...700.1716W,2015ApJ...805..114W, 2012SCPMA..55.1316M,2012ApJ...753...52L,2012ApJ...745L..18A}). In addition to the density distribution, we also examine the distributions of gas pressure $p$, horizontal velocity component $v_{x}$, and vertical velocity component $v_{y}$ (see rows 2 to 4 in Figure \ref{fig:evolution}). We notice that, comparing with the distribution of the density, those of the velocity and the pressure reveal more fine structures. These features may not be easily recognized in images that basically display the brightness of targets. If the information about the Doppler shift could be obtained, say via spectroscopic approach, they may become apparent.

At $t = 33$ s, as shown in the second column of Figure \ref{fig:evolution}, two semicircular wave fronts (``w${1}$" and ``w${2}$") are propagating forward of the flux rope. In contour plots for velocity (Figure \ref{fig:evolution}(e) and \ref{fig:evolution}(h)), the wave patterns are quite rich and impressive. Previous numerical experiments (e.g., see also \citealt{2019MNRAS.490.2918X,2009ApJ...700.1716W,2015ApJ...805..114W, 2012SCPMA..55.1316M}) usually displayed only one wave front, while the eruption process investigated in the present work produces three sets of wave fronts, which can even be clearly distinguished in the density distribution plot (Figure \ref{fig:evolution}(c)).

In general, the generated waves seem to propagate isotropically, i.e., the local propagation speed does not seem to depend on the direction. This is governed by the fact that the fast speed is almost uniform in the region near the flux as shown in Figure \ref{fig:initial}(b). These wave fronts appear strikingly similar to the quasi-periodic fast magnetoacoustic wave trains displayed by \cite{2012ApJ...753...52L}. We shall discuss these results later in more detail and compare them with observation. Furthermore, density contours in Figure \ref{fig:evolution}(a) through Figure \ref{fig:evolution}(c) showed that the wavefront at the low altitude is more significant than that at the high altitude . This difference may be attributed to the gravitationally stratified atmosphere, in which the plasma is dense at low altitudes and tenuous at high altitudes, therefore, the response of low atmosphere to the wave is more apparent than that of high atmosphere.

\subsection{Characteristics of Wavefronts }

\begin{figure}[t]
\centerline{\includegraphics[width=1\textwidth,clip=]{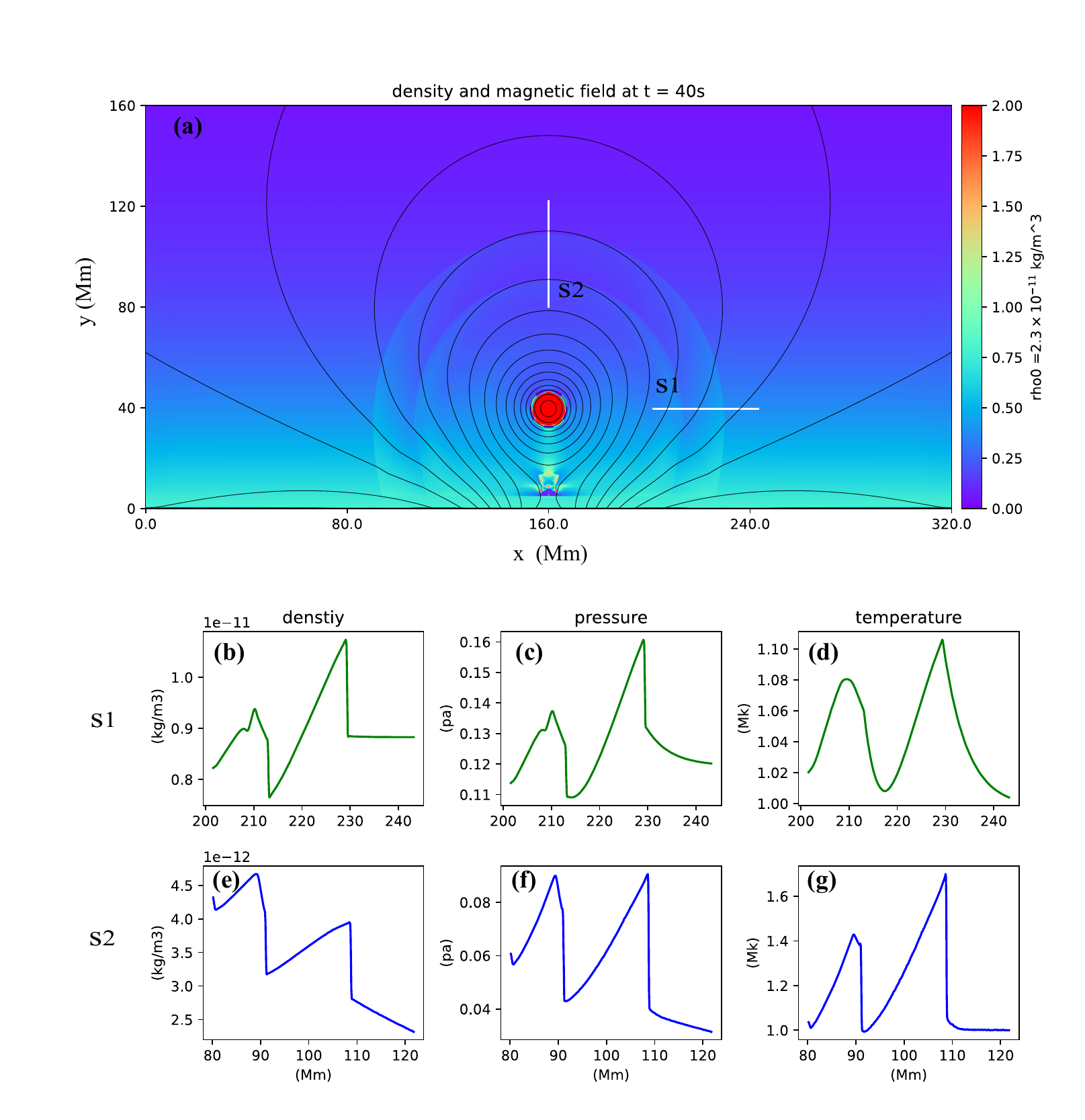}}
\caption{Physical quantities variations in slices crossing wavefronts. (a) displays a snapshot of density and magnetic filed at t = 40 s. Magnetic field lines are depicted by solid gray lines. (b-d) and (e-g) illustrate changes in density, pressure and temperature along slices s1 and s2, respectively.}
\label{fig:two_shock}
\end{figure}

According to Figures \ref{fig:evolution}(a)-\ref{fig:evolution}(c), these fronts appear sequentially in roughly symmetric shapes ahead of the flux rope. To look into their physical properties, we further analyze their apparent behaviors. Figure \ref{fig:two_shock}(a) displays the density distribution at $t = 40$~s, with black solid lines for magnetic field lines. From the figure, we see that these two wavefronts propagate successively on the solar surface to a distance of a few $10^{4}$~km, and an obvious deflection of the magnetic field occurs on the wavefront with an apparent enhancement in the tangential component of the magnetic field. This suggests that these waves are the fast-mode shock.

To quantitatively reveal the property of these fronts, we analyze the changes in the relevant parameters before and after sweeping of the front along two white lines, s${1}$ and s${2}$. Here line s${1}$ crosses the front horizontally, with the upstream (undisturbed medium) on the right side of the  front and the downstream (disturbed medium) on the left side, s${2}$ crosses the shock front vertically, with the upstream on the top side and the downstream on the bottom side of the front. Figures \ref{fig:two_shock}(b)-\ref{fig:two_shock}d show the distribution of density, pressure, and temperature along s${1}$, while Figures \ref{fig:two_shock}(e)-\ref{fig:two_shock}(g) show the distributions along s${2}$.

 We found that the density and the pressure along s${1}$ exhibit clear discontinuities, while the temperature change is relatively smooth. After the passage of the first wavefront, the plasma density, pressure, and temperature increase by 1.25, 1.23, and 1.1 times, respectively; after the passage of the second wave, the increases become 1.17, 1.18, and 1.08 times, respectively. This indicates that the first wavefront is stronger and impacts the environment more significantly than the second one. The distributions of the above parameters along s${2}$ are similar to those along s${1}$, and also exhibit clear discontinuity layers in density and pressure. After the passage of the first wavefront, the plasma density, pressure, and temperature increase by 1.48, 2.25, and 1.7 times, respectively; and on the second wavefront, the corresponding changes become 1.48, 2.14, and 1.41 times, respectively. In addition, we also noticed that the properties of the plasma and magnetic field along s${1}$ and s${2}$ show slight differences because of the effect of gravitational stratification. Furthermore, apparent similarities could be recognized when comparing the magnetoacoustic shock structure shown in Figure \ref{fig:two_shock} with the formation of the shock as a result of the leakage of fast-mode waves from a plasma slab studied by \cite{2017ApJ...847L..21P}. In summary, the observed discontinuity layers along s${1}$ and s${2}$ suggest again that these disturbances are fast-mode shock.

\begin{figure}[t]
\centerline{\includegraphics[width=1\textwidth,clip=]{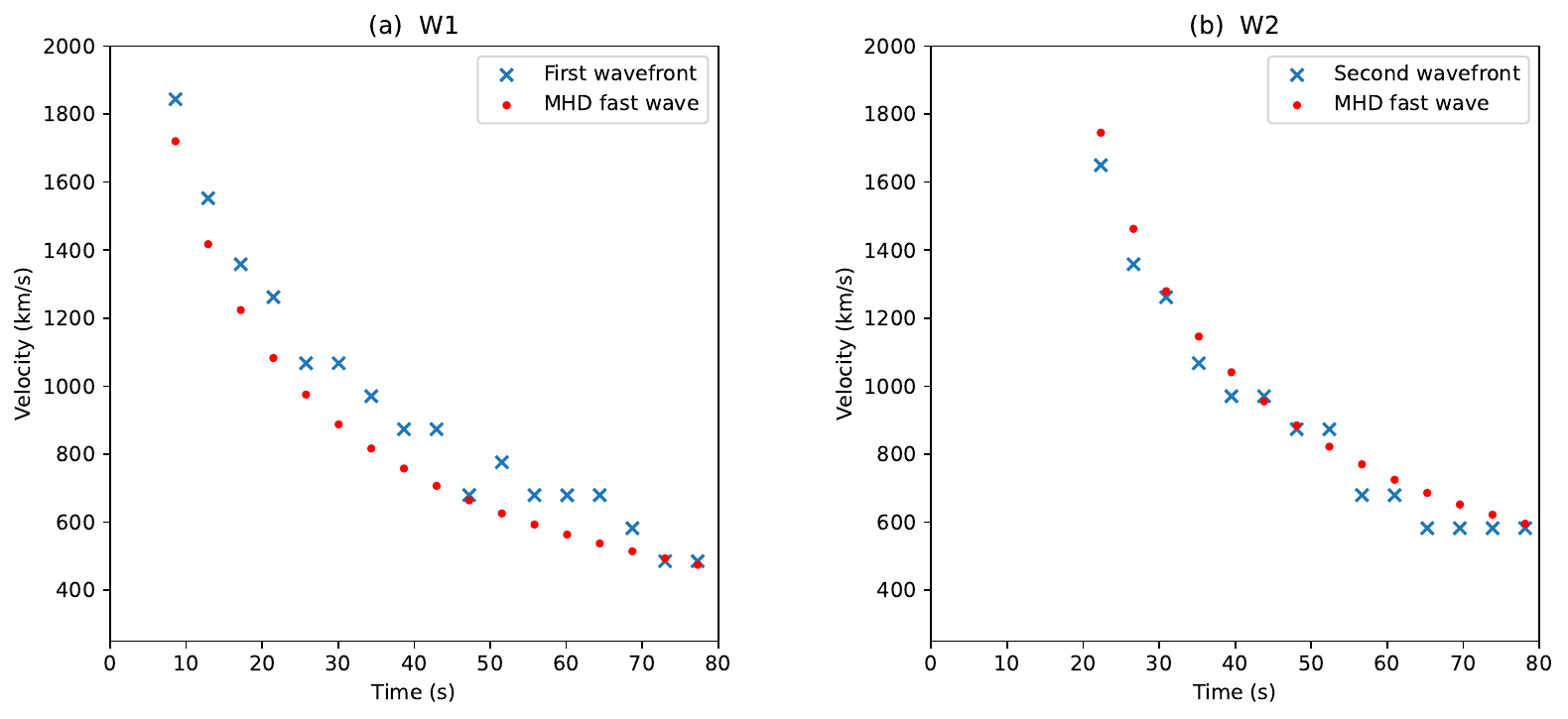}}
\caption{Propagation speeds for the first wavefront ``w1"  in (a) and the second wavefront ``w2"  in (b) versus time. The blue symbols ``x" are for the velocities of the wavefronts, while the red dots for the local fast speed that could be deduced from Figure \ref{fig:initial}(b) correspondingly.}
\label{fig:wavespeed}
\end{figure}

To further confirm the fast-mode properties of these perturbations, we compare their velocities with the local fast magnetosonic wave speeds. Figures \ref{fig:wavespeed}(a) and \ref{fig:wavespeed}(b) show the variations of the velocity versus time for the first and second  wavefronts, respectively. The blue symbols ``x" are for the velocities of the wavefronts, while the red dots for the local fast speed that could be deduced from Figure \ref{fig:initial}(b) accordingly. In general, as the propagation distance increases, the velocities of the wavefronts decrease. From Figure \ref{fig:wavespeed}(a), we can see that in the early stages, the velocity of the first  wavefront is greater than the local fast magnetosonic wave speed, with an average Mach number of 1.12 (relative to the fast magnetosonic wave speed); on the other hand, as the wave further propagates, it gradually degrades into the fast magnetosonic wave. As for the second wavefront, its initial velocity (approximately 1700 ~km~s$^{-1}$) is lower than that of the first wavefront (approximately 1900 ~km~s$^{-1}$), and its velocity is approximately equal to the fast magnetosonic wave speed.

\begin{figure}[t]
\centerline{\includegraphics[width=1\textwidth,clip=]{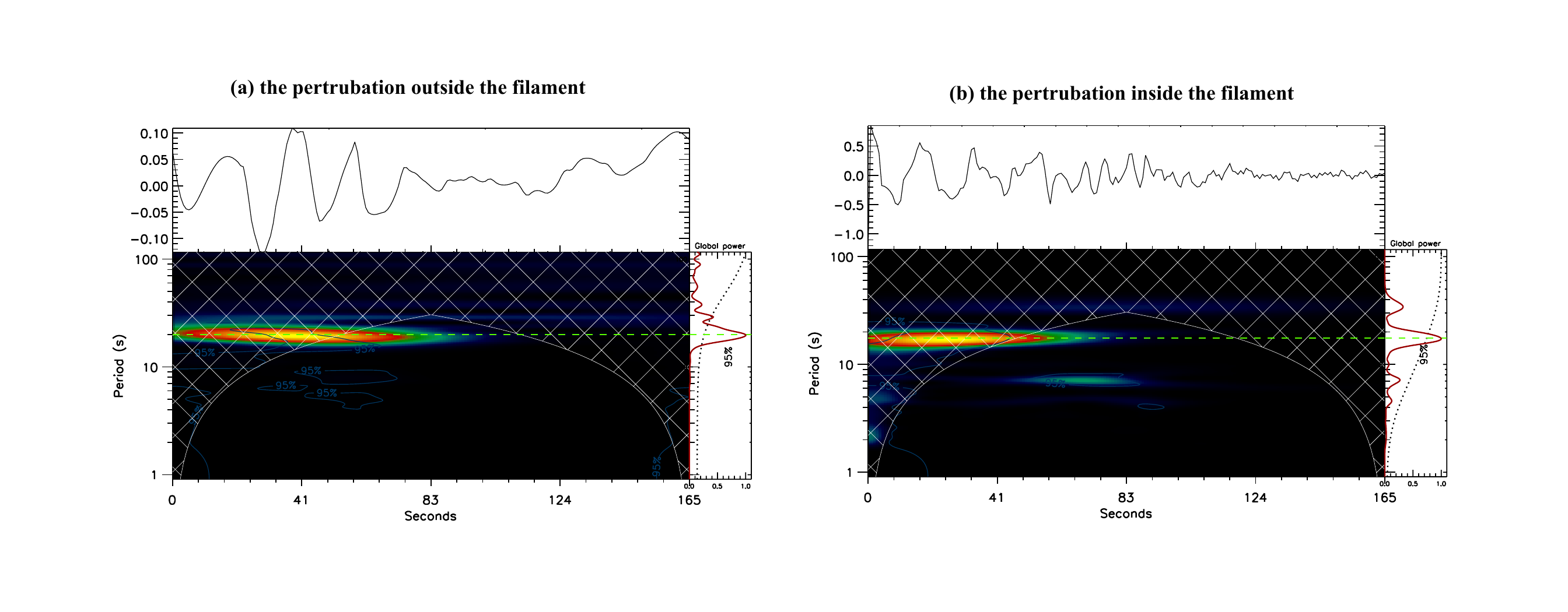}}
\caption{Periodicity analysis of quasi-periodic wave trains.
(a) denotes the periodicity analysis of the fixed point outside the flux rope (marked as ``A" in Fig \ref{fig:initial}), and   (b) denotes that of the central point of the flux rope.}
\label{fig:period}
\end{figure}

Moreover, the wave patterns shown in Figure \ref{fig:evolution} are suggestive of the periodical generation of these waves. To investigate this process, we conducted a wavelet analysis of the energy variation versus time at a specific location outside the flux rope (labeled as ``A" in Figure \ref{fig:initial}). The results are presented in Figure \ref{fig:period}(a), revealing a disturbance signal with a period of approximately 20 seconds. Simultaneously, we also performed a similar analysis of the energy disturbance at the center of the flux rope, and the results are displayed in Figure \ref{fig:period}(b). Interestingly, we found that the internal disturbance also exhibits clear periodical feature, and the period is roughly the same as that of the external disturbance. This suggests the connection of the origins of the two disturbances to each other, which we will further discuss below.

\subsection{Origin of  Wave Trains}

\begin{figure}[t]
\begin{interactive}{animation}{form_process.mp4}
\centerline{\includegraphics[width=0.7\textwidth,clip=]{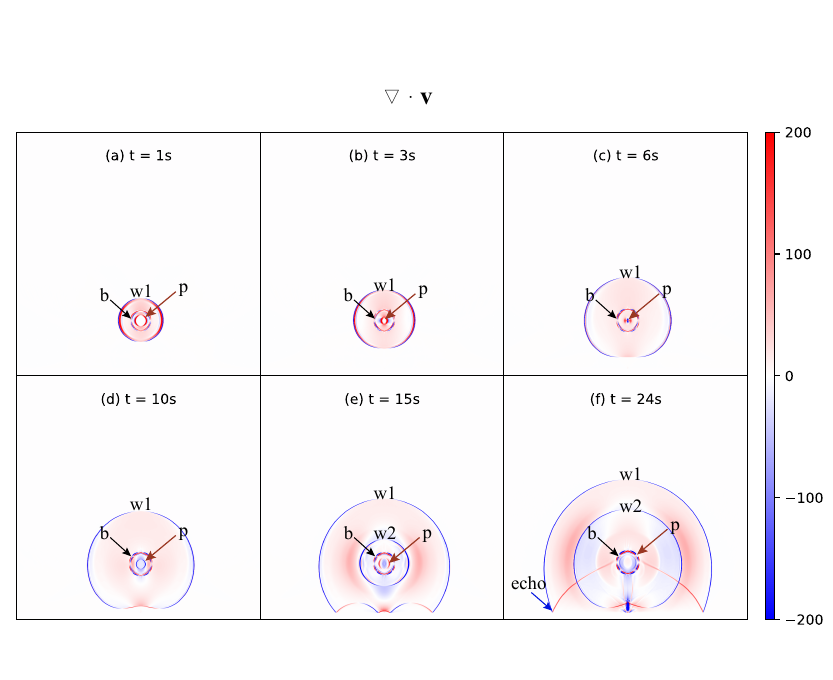}}
\end{interactive}
\caption{The velocity divergence at different times. The red arrow ``p'' is the inner perturbation, and the black arrow ``b'' the boundary of the flux rope. ``w${1}$" and ``w${2}$" are the first and the second wavefronts, respectively. The animated velocity divergence images display the propagation of internal disturbances and the formation of QFP waves in the time interval from 0 s to 50.67 s with a duration of 2 s. Static image only presents information at a  few key moments, while the animation shows the evolution. An animated version of this figure is available.}
\label{fig:origin}
\end{figure}

We calculate the velocity divergence at different times (Figure \ref{fig:origin} and the supplementary movie) to analyze the origin and property of disturbances. According to Equation (\ref{continum}), the velocity divergence describes the compression ($\nabla \cdot \bf{v} < 0$) and expansion ($\nabla \cdot \bf{v} > 0$) of the plasma, making it an effective indicator of the shock.

As mentioned earlier, in the numerical experiments, the initial position of the flux rope is close to the critical point, so it stays in this position for a short period before losing the equilibrium and then rapidly moving upward after losing the equilibrium. Simultaneously, the internal equilibrium of the flux rope is also disrupted. Because of the compressibility of the plasma, once a disturbance occurs, it propagates outward in the form of the wave. With losing the internal equilibrium, two waves are invoked inside the flux rope, and propagate inward and outward, respectively. Some of the outward-propagating waves eventually leave the flux rope and enter the surrounding corona, while other part is reflected at the edge of the flux rope, forming inward-propagating waves. The inward waves eventually converge at the center of the flux rope and then rebound outward. As a result, we can see an image of the flux rope constantly oscillating and waves continuously propagating outward from its edges. The oscillations of the flux rope eventually dies down as a result of the dissipation and the wave leakage into the external medium.

Figure \ref{fig:origin} qualitatively describes such a process. At $t=1$~s (see Figure \ref{fig:origin}(a)), an outward-propagating wave (labeled as ``w${1}$'') is observed outside the flux rope, corresponding to the first wave shown in Figure \ref{fig:evolution}. It is caused by the loss of equilibrium of the flux rope and its rapid upward motion. Meanwhile, a circular inward-propagating disturbance (labeled as ``p") appears inside the flux rope. Between w${1}$ and p, a circular boundary interface exists and does not change significantly with time, which is actually the boundary of the flux rope (labeled as ``b"). From Figure \ref{fig:origin}(b) and \ref{fig:origin}(c), we realize that w$_{1}$ continues to propagate outward, while p contracts further toward the center of the flux rope until it shrinks to a point.

Later, at $t=10$~s  (see Figure \ref{fig:origin}(d)), the downward portion of w${1}$ has propagated into the lower solar atmosphere, while p starts expanding outward from the center of the flux rope. At $t=15$~s (see Figure \ref{fig:origin}(e)), the downward portion of w${1}$ produces aforementioned echoes. Simultaneously, disturbance p encounters boundary b, and a portion of it transmits through b and propagates outward, forming the second wavefront (labeled as ``w2" in Figure \ref{fig:origin}(e)), which is consistent with the results of \cite{2017ApJ...847L..21P}. Another portion reflects and propagates back toward the center of the flux rope again. The propagation of p as described above forms a cycle that continuously triggers oscillations inside the flux rope, leaking waves on its surface until the extra energy inside the flux rope totally dissipates. This scenario well displays the creation and propagation of various waves/shocks around the flux rope during the eruption, which could find observational counterparts easily.

\section{Observational consequences}\label{part4}

\begin{figure}[t]
\centerline{\includegraphics[width=0.9\textwidth,clip=]{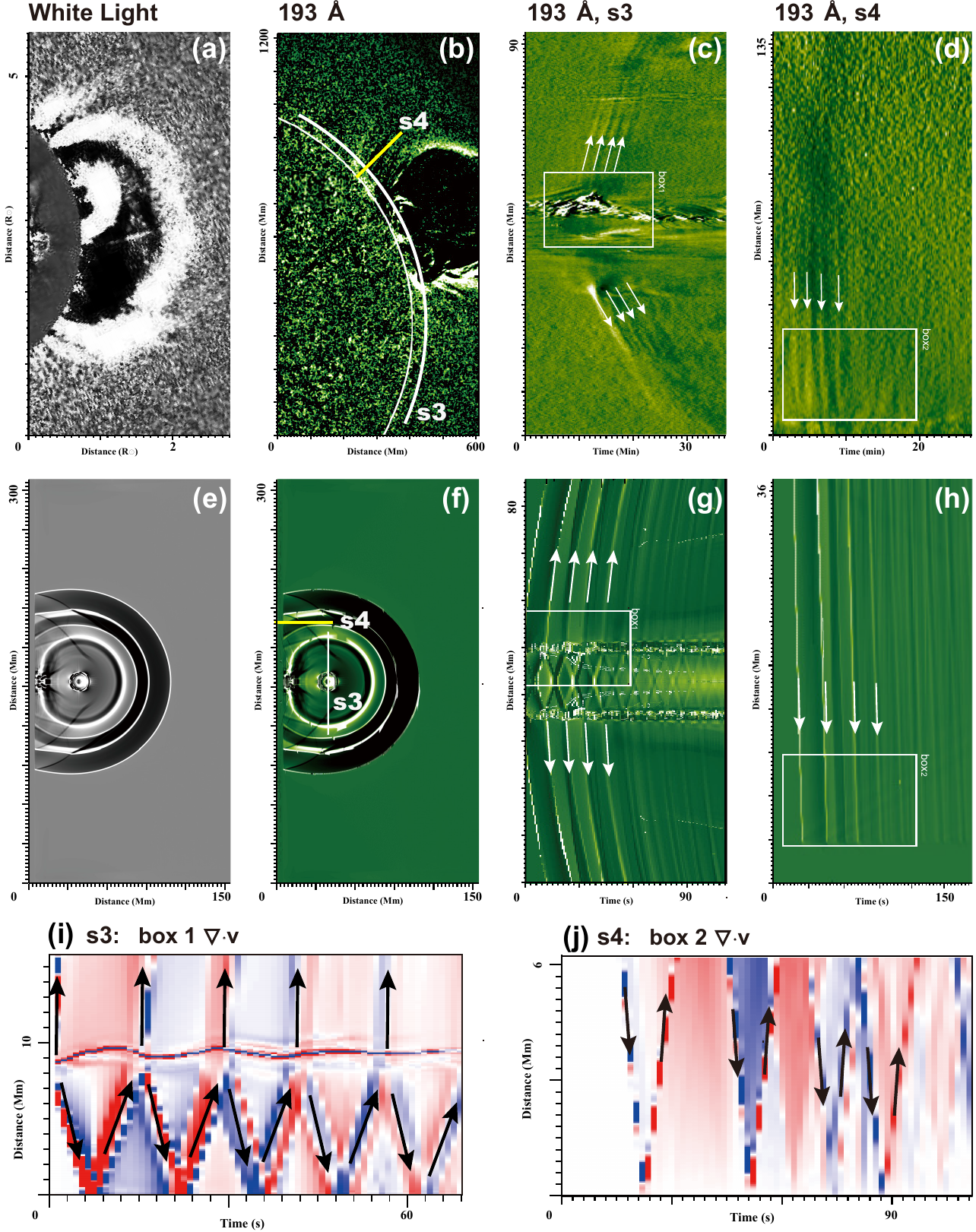}}
\caption{Comparison between observations and the simulation. The top row displays the observations while the middle row the simulation. Specifically, (a) depicts an event that occurred on May 7, 2012, where multiple wavefronts were detected on both sides of a flux rope through the white-light coronagraph COR1 (\citealt{2017ApJ...844..149K}). (b) shows an event that occurred on September 8-9, 2010, where QFP waves were observed in the AIA image at 193 \AA~ (\citealt{2012ApJ...753...52L}). (c) and (d) are the time-sequences of the brightness distribution along curve s${3}$ and line s${4}$ in (b). (i) and (j) are magnified images of the velocity divergence within the boxed regions of (g) and (h). }
\label{fig:euvsyns}
\end{figure}

The quasi-periodic fast-mode magnetoacoustic waves propagating on the solar surface triggered by CME have been reported by a few authors. \cite{2017ApJ...844..149K} reported several wavefronts around the flux rope observed by the white-light coronagraph on board STERO-A (Figure \ref{fig:euvsyns}(a)).  \cite{2012ApJ...753...52L} performed detailed studies of the similar phenomenon observed in EUV by SDO/AIA (Figure \ref{fig:euvsyns}(b)).

Figures \ref{fig:euvsyns}(c) and \ref{fig:euvsyns}(d) show the time-sequences of the brightness distributions along curve s${3}$ and line s${4}$ specified in Figure \ref{fig:euvsyns}(b), revealing information about the propagation of the wavefronts in various directions. Figure \ref{fig:euvsyns}(c) demonstrated multiple wavefronts (indicated by white arrows) are successively ejected from the boundary of CME core. Box1 in the figure highlights how the upper part of the wavefront leaks out. Figure \ref{fig:euvsyns}(d) describes the situation in the vertical direction along s${4}$, where multiple downward wavefronts (indicated by white arrows) reach the solar surface, and the echo could be recognized. Box2 in the figure outlines the region where the waves interact with the lower solar atmosphere. Interested readers refer to \cite{2019MNRAS.490.2918X} for more information regarding the identification and analysis of the echo on the solar surface.

To compare with observations, we create synthetic images in white-light and EUV wavelength on the base of the simulation results. A synthetic images is basically constructed according to instrument response functions and atomic databases. The instrument response function provides the response of the detector in a given wavelength to the plasma radiation at a given temperature. The detector responds to the radiation in the way of:
\begin{equation}
R_{f} = \int n_{e}^{2} f_{i}(T_{e},n_{e})dl. \qquad (DN/s) \label{synthetic}
\end{equation}
Here, $dl$ is the length element along the line of sight (LOS), $DN$  the number of electrons produced by the photons received by the instrument, $n_{e}$ the electron density, and $f_{i}(T_{e},n_{e})$ the response function of the instrument for density $n_{e}$ and temperature $T_{e}$. A synthetic image then could be obtained provided the form of $f_{i}(T_{e}, n_{e})$, and distributions of $T_{e}$ as well as $n_{e}$ in LOS are known.

Figure \ref{fig:euvsyns}(e) shows the synthetic image in white-light, and  Figures \ref{fig:euvsyns}(f) through \ref{fig:euvsyns}(h) provide the information in 193~\AA. Comparing observations (Figures \ref{fig:euvsyns}(a) and \ref{fig:euvsyns}(b)) with the synthetic images (Figures \ref{fig:euvsyns}(e) and \ref{fig:euvsyns}(f)), we find that the simulation results exhibit multiple wavefronts in both wavelengths, but observations display little bit different scenario such that these waves are almost invisible in front of CME, but apparent on the side of CME. This could be due to the fact that the density on the CME side is higher than  that in front of CME, and the disturbance on the CME side may produce more apparent observational consequences; in the numerical experiments, on the other hand, manifestations of the synthetic image could be manipulated by adjusting the range of color bar.

To further justify that the multiple wavefront scenario is the oscillation inside the flux rope in origin, we deduce the time-sequences of the brightness distributions in directions parallel and perpendicular to the solar surface as indicated by Figure \ref{fig:euvsyns}(f). The results are displayed in Figures \ref{fig:euvsyns}(g) and \ref{fig:euvsyns}(h). Comparing Figure \ref{fig:euvsyns}(c) with Figure \ref{fig:euvsyns}(g), we see several wave fronts in both observational and synthetic images, and these wavefront structures originate from the core region of the eruption. Furthermore, our simulation, on which the synthetic images are created,  definitely confirms that these wavefronts  result from leaking of the oscillation inside the flux rope.

In addition, a set of X-like patterns inside the flux rope are observed. In Figure \ref{fig:euvsyns}(i), we plot velocity divergence in the region inside box1 in Figure \ref{fig:euvsyns}(g). The oscillating features in the radial direction of the flux rope could be recognized, and the leakage of the oscillation through the boundary of the flux rope is also seen clearly. In this process, the waves associated the oscillation propagate both inward and outward inside the flux rope. Concentrating and bouncing of the the inward wave at the center of the flux rope constitute the X-shape feature as shown by  shown in Figure \ref{fig:euvsyns}(g) and \cite{2013A&A...560A..97P}.

Comparing Figure \ref{fig:euvsyns}(d) with Figure \ref{fig:euvsyns}(h), we find multiple wavefronts propagating downward in both observations and synthetic images. To study the interaction of the downward wavefronts with the solar surface, we calculate the corresponding velocity divergence in the region indicated by box2 in Figure \ref{fig:euvsyns}(h), and the result is plotted in Figure \ref{fig:euvsyns}(j), which shows clearly the pattern of the echo. The similar result was also reported by \cite{2019MNRAS.490.2918X}, with the only difference being that \cite{2019MNRAS.490.2918X} showed a single echo, whereas several echos could be identified in Figure \ref{fig:euvsyns}(j).

\section{Summary And Discussions} \label{part5}

On the basis of the analytical model of \cite{1993ApJ...417..368I}, we investigated the evolution in a magnetic configuration that includes an electric current-carrying  flux rope and possesses a relatively simple background magnetic field. Our results confirmed that the coronal magnetic structure described by the model can quasi-statically evolve through a set of configurations in the stable equilibrium to an unstable configuration, eventually leading to an eruption in a catastrophic fashion. Comparing with the works by \cite{2001JGR...10625053L}, \cite{2016A&A...590A.120K, 2018JASTP.172...40K}, as well as \cite{2022ApJ...933..148C}, we notice that the evolutionary behaviors of the magnetic structure studied here is simple and straightforward such that the displacement of the flux rope only moves in the $y$-direction during both quasi-static and dynamic stages. \cite{2001JGR...10625053L},
\cite{2016A&A...590A.120K, 2018JASTP.172...40K}, and \cite{2022ApJ...933..148C} found that the motion of the flux rope could be in both $x$- and $y$-directions(here the $y$-direction is the vertical direction) as the background field is complex, and even the loss of equilibrium in the flux rope may not be able to develop to a plausible eruption eventually. This might be more realistic.

However, the purpose of this work is to look into the origin of the QFP wave during the eruption no matter how the eruption is triggered. So a simpler magnetic configuration was chosen and the straightforward evolutionary behavior in the configuration was studied. After confirming the occurrence of the loss of equilibrium of the flux rope, we subsequently investigated the occurrence of QFP waves in the solar eruption. Our results showed that arc-like wavefronts successively appear on the surface of the flux rope and propagate outward at speed of a few $10^{3}$~km~s$^{-1}$ as the flux rope moves out.

We analyzed the mechanism and propagating features of these wavefronts , and found out they were moving at speeds faster or marginally faster than the local fast magneto-acoustic wave speed, which implies their nature of the fast-mode shock. Looking into the velocity divergence of the wavefront, we discovered that these QFP waves originate from disturbances within the flux rope as a result of the loss of the internal equilibrium, and part of the disturbance leaks through the flux rope boundary out to the surrounding corona, forming the QFP waves observed. When QFP waves reach the lower solar atmosphere, they generate multiple echos. In summary, the role of flux rope in the origin of the wave trains observed in the solar eruption is twofold. First, the loss of the internal equilibrium of the flux rope ignites oscillations inside the flux rope; and second, the flux rope behaves like a leaking wave guide from which the oscillation successively leaks out to the surrounding corona, invoking the wave train with period of several tens of second. The numerical simulation of \cite{2021ApJ...911L...8W} showed that the period of the wave train is around 30~s.

In addition to the above point regarding the origin of wave trains, \cite{2021ApJ...911L...8W}  suggested that the wave train is a component of the large-scale EUV waves. But this mechanism is hard to account for the fact that various wave fronts come from roughly the same region, and that velocities of different wave fronts are approximately the same.

Numerical experiments performed here duplicated the large-scale QFP waves observed in the lower solar corona by \cite{2012ApJ...753...52L} as well. In addition, \cite{2017ApJ...844..149K} also identified QFP waves in the high corona on the basis of analyzing the data from the white-light coronagraph COR1 on STEREO-A. To our knowledge, no author reports simultaneous identification of QFP waves in both the low and the high solar corona regions in a single eruptive event. Furthermore,  because of the limit to the computational resources, our numerical experiments could not investigate QFP in both regions simultaneously, either; so it is very hard to reveal the relations of the wave train in the low to that in the high corona for the time being. But the possibility exists that the QFP wave in the low corona propagates upward, reaches to the large altitudes, and evolves into that in the high corona.

Overall, to definitively determine the mechanisms for their origins, it is necessary to identify and compare the processes of QFP wave generation and propagation in both the low and higher solar corona regions. This requires observational instruments with large FOV that cover both the low and the high corona regions with sufficiently high time cadence. The existing observational data may not fully meet these requirements, we hope that advanced technology and observational instruments in the future would be able to help solve this puzzle.

\vspace{2em}
\noindent
This work was supported by the National Key R\&D Program of China No.2022YFF0503804,  the NSFC grants 11933009, 12273107, U2031141, and 12073073, grants associated with the Yunling Scholar Project of the Yunnan Province, the Yunnan Province Scientist Workshop of Solar Physics, the Applied Basic Research of Yunnan Province (2019FB005, 202101AT070018), Yunnan Key Laboratory of Solar Physics and Space Exploration of Code 202205AG070009. JY and ZM. also acknowledge the support by grants associated with the Yunnan Revitalization Talent Support Program, the Foundation of the Chinese Academy of Sciences (Light of West China Program). The numerical computation in this paper was carried out on the computing facilities of the Computational Solar Physics Laboratory of Yunnan Observatories (CosPLYO).

\bibliography{hjl}{}

\begin{thebibliography}{}
\expandafter\ifx\csname natexlab\endcsname\relax\def\natexlab#1{#1}\fi
\providecommand{\url}[1]{\href{#1}{#1}}
\providecommand{\dodoi}[1]{doi:~\href{http://doi.org/#1}{\nolinkurl{#1}}}
\providecommand{\doeprint}[1]{\href{http://ascl.net/#1}{\nolinkurl{http://ascl.net/#1}}}
\providecommand{\doarXiv}[1]{\href{https://arxiv.org/abs/#1}{\nolinkurl{https://arxiv.org/abs/#1}}}

\bibitem[{{Asai} {et~al.}(2012){Asai}, {Ishii}, {Isobe}, {Kitai}, {Ichimoto},
  {UeNo}, {Nagata}, {Morita}, {Nishida}, {Shiota}, {Oi}, {Akioka}, \&
  {Shibata}}]{2012ApJ...745L..18A}
{Asai}, A., {Ishii}, T.~T., {Isobe}, H., {et~al.} 2012, \apjl, 745, L18,
  \dodoi{10.1088/2041-8205/745/2/L18}

\bibitem[{{Banerjee} {et~al.}(2021){Banerjee}, {Krishna Prasad}, {Pant},
  {McLaughlin}, {Antolin}, {Magyar}, {Ofman}, {Tian}, {Van Doorsselaere}, {De
  Moortel}, \& {Wang}}]{2021SSRv..217...76B}
{Banerjee}, D., {Krishna Prasad}, S., {Pant}, V., {et~al.} 2021, \ssr, 217, 76,
  \dodoi{10.1007/s11214-021-00849-0}

\bibitem[{{Cally}(1986)}]{1986SoPh..103..277C}
{Cally}, P.~S. 1986, \solphys, 103, 277, \dodoi{10.1007/BF00147830}

\bibitem[{{Chen} {et~al.}(2022){Chen}, {Ye}, {Mei}, {Shen}, {Roussev},
  {Forbes}, {Lin}, \& {Ziegler}}]{2022ApJ...933..148C}
{Chen}, Y., {Ye}, J., {Mei}, Z., {et~al.} 2022, \apj, 933, 148,
  \dodoi{10.3847/1538-4357/ac73ef}

\bibitem[{{Cooper} {et~al.}(2003){Cooper}, {Nakariakov}, \&
  {Williams}}]{2003A&A...409..325C}
{Cooper}, F.~C., {Nakariakov}, V.~M., \& {Williams}, D.~R. 2003, \aap, 409,
  325, \dodoi{10.1051/0004-6361:20031071}

\bibitem[{{De Moortel} {et~al.}(2000){De Moortel}, {Ireland}, \&
  {Walsh}}]{2000A&A...355L..23D}
{De Moortel}, I., {Ireland}, J., \& {Walsh}, R.~W. 2000, \aap, 355, L23

\bibitem[{{Forbes}(1990)}]{1990JGR....9511919F}
{Forbes}, T.~G. 1990, \jgr, 95, 11919, \dodoi{10.1029/JA095iA08p11919}

\bibitem[{{Gopalswamy} {et~al.}(2009){Gopalswamy}, {Yashiro}, {Temmer},
  {Davila}, {Thompson}, {Jones}, {McAteer}, {Wuelser}, {Freeland}, \&
  {Howard}}]{2009ApJ...691L.123G}
{Gopalswamy}, N., {Yashiro}, S., {Temmer}, M., {et~al.} 2009, \apjl, 691, L123,
  \dodoi{10.1088/0004-637X/691/2/L123}

\bibitem[{{Isenberg} {et~al.}(1993){Isenberg}, {Forbes}, \&
  {Demoulin}}]{1993ApJ...417..368I}
{Isenberg}, P.~A., {Forbes}, T.~G., \& {Demoulin}, P. 1993, \apj, 417, 368,
  \dodoi{10.1086/173319}

\bibitem[{{Keppens} {et~al.}(2012){Keppens}, {Meliani}, {van Marle}, {Delmont},
  {Vlasis}, \& {van der Holst}}]{2012JCoPh.231..718K}
{Keppens}, R., {Meliani}, Z., {van Marle}, A.~J., {et~al.} 2012, Journal of
  Computational Physics, 231, 718, \dodoi{10.1016/j.jcp.2011.01.020}

\bibitem[{{Kolotkov} {et~al.}(2016){Kolotkov}, {Nistic{\`o}}, \&
  {Nakariakov}}]{2016A&A...590A.120K}
{Kolotkov}, D.~Y., {Nistic{\`o}}, G., \& {Nakariakov}, V.~M. 2016, \aap, 590,
  A120, \dodoi{10.1051/0004-6361/201628501}

\bibitem[{{Kolotkov} {et~al.}(2018){Kolotkov}, {Nistic{\`o}}, {Rowlands}, \&
  {Nakariakov}}]{2018JASTP.172...40K}
{Kolotkov}, D.~Y., {Nistic{\`o}}, G., {Rowlands}, G., \& {Nakariakov}, V.~M.
  2018, Journal of Atmospheric and Solar-Terrestrial Physics, 172, 40,
  \dodoi{10.1016/j.jastp.2018.03.005}

\bibitem[{{Kumar} {et~al.}(2017){Kumar}, {Nakariakov}, \&
  {Cho}}]{2017ApJ...844..149K}
{Kumar}, P., {Nakariakov}, V.~M., \& {Cho}, K.-S. 2017, \apj, 844, 149,
  \dodoi{10.3847/1538-4357/aa7d53}

\bibitem[{{Lemen} {et~al.}(2012){Lemen}, {Title}, {Akin}, {Boerner}, {Chou},
  {Drake}, {Duncan}, {Edwards}, {Friedlaender}, {Heyman}, {Hurlburt}, {Katz},
  {Kushner}, {Levay}, {Lindgren}, {Mathur}, {McFeaters}, {Mitchell}, {Rehse},
  {Schrijver}, {Springer}, {Stern}, {Tarbell}, {Wuelser}, {Wolfson}, {Yanari},
  {Bookbinder}, {Cheimets}, {Caldwell}, {Deluca}, {Gates}, {Golub}, {Park},
  {Podgorski}, {Bush}, {Scherrer}, {Gummin}, {Smith}, {Auker}, {Jerram},
  {Pool}, {Soufli}, {Windt}, {Beardsley}, {Clapp}, {Lang}, \&
  {Waltham}}]{2012SoPh..275...17L}
{Lemen}, J.~R., {Title}, A.~M., {Akin}, D.~J., {et~al.} 2012, \solphys, 275,
  17, \dodoi{10.1007/s11207-011-9776-8}

\bibitem[{{Li} {et~al.}(2020){Li}, {Antolin}, {Guo}, {Kuznetsov}, {Pascoe},
  {Van Doorsselaere}, \& {Vasheghani Farahani}}]{2020SSRv..216..136L}
{Li}, B., {Antolin}, P., {Guo}, M.~Z., {et~al.} 2020, \ssr, 216, 136,
  \dodoi{10.1007/s11214-020-00761-z}

\bibitem[{{Lin} {et~al.}(2001){Lin}, {Forbes}, \&
  {Isenberg}}]{2001JGR...10625053L}
{Lin}, J., {Forbes}, T.~G., \& {Isenberg}, P.~A. 2001, \jgr, 106, 25053,
  \dodoi{10.1029/2001JA000046}

\bibitem[{{Lin} {et~al.}(2010){Lin}, {Shen}, \& {Wang}}]{2010cosp...38.1800L}
{Lin}, J., {Shen}, C., \& {Wang}, H. 2010, in 38th COSPAR Scientific Assembly,
  Vol.~38, 4

\bibitem[{{Liu} {et~al.}(2012){Liu}, {Ofman}, {Nitta}, {Aschwanden},
  {Schrijver}, {Title}, \& {Tarbell}}]{2012ApJ...753...52L}
{Liu}, W., {Ofman}, L., {Nitta}, N.~V., {et~al.} 2012, \apj, 753, 52,
  \dodoi{10.1088/0004-637X/753/1/52}

\bibitem[{{Liu} {et~al.}(2011){Liu}, {Title}, {Zhao}, {Ofman}, {Schrijver},
  {Aschwanden}, {De Pontieu}, \& {Tarbell}}]{2011ApJ...736L..13L}
{Liu}, W., {Title}, A.~M., {Zhao}, J., {et~al.} 2011, \apjl, 736, L13,
  \dodoi{10.1088/2041-8205/736/1/L13}

\bibitem[{{McLaughlin} {et~al.}(2018){McLaughlin}, {Nakariakov}, {Dominique},
  {Jel{\'\i}nek}, \& {Takasao}}]{2018SSRv..214...45M}
{McLaughlin}, J.~A., {Nakariakov}, V.~M., {Dominique}, M., {Jel{\'\i}nek}, P.,
  \& {Takasao}, S. 2018, \ssr, 214, 45, \dodoi{10.1007/s11214-018-0478-5}

\bibitem[{{Mei} {et~al.}(2012{\natexlab{a}}){Mei}, {Shen}, {Wu}, {Lin},
  {Murphy}, \& {Roussev}}]{2012MNRAS.425.2824M}
{Mei}, Z., {Shen}, C., {Wu}, N., {et~al.} 2012{\natexlab{a}}, \mnras, 425,
  2824, \dodoi{10.1111/j.1365-2966.2012.21625.x}

\bibitem[{{Mei} {et~al.}(2012{\natexlab{b}}){Mei}, {Udo}, \&
  {Lin}}]{2012SCPMA..55.1316M}
{Mei}, Z., {Udo}, Z., \& {Lin}, J. 2012{\natexlab{b}}, Science China Physics,
  Mechanics, and Astronomy, 55, 1316, \dodoi{10.1007/s11433-012-4752-3}

\bibitem[{{Miao} {et~al.}(2019){Miao}, {Liu}, {Shen}, {Li}, {Abidin},
  {Elmhamdi}, \& {Kordi}}]{2019ApJ...871L...2M}
{Miao}, Y.~H., {Liu}, Y., {Shen}, Y.~D., {et~al.} 2019, \apjl, 871, L2,
  \dodoi{10.3847/2041-8213/aafaf9}

\bibitem[{{Nakariakov} {et~al.}(2004){Nakariakov}, {Arber}, {Ault},
  {Katsiyannis}, {Williams}, \& {Keenan}}]{2004MNRAS.349..705N}
{Nakariakov}, V.~M., {Arber}, T.~D., {Ault}, C.~E., {et~al.} 2004, \mnras, 349,
  705, \dodoi{10.1111/j.1365-2966.2004.07537.x}

\bibitem[{{Nakariakov} {et~al.}(2012){Nakariakov}, {Hornsey}, \&
  {Melnikov}}]{2012ApJ...761..134N}
{Nakariakov}, V.~M., {Hornsey}, C., \& {Melnikov}, V.~F. 2012, \apj, 761, 134,
  \dodoi{10.1088/0004-637X/761/2/134}

\bibitem[{{Nakariakov} \& {Kolotkov}(2020)}]{2020ARA&A..58..441N}
{Nakariakov}, V.~M., \& {Kolotkov}, D.~Y. 2020, \araa, 58, 441,
  \dodoi{10.1146/annurev-astro-032320-042940}

\bibitem[{{Nakariakov} \& {Melnikov}(2009)}]{2009SSRv..149..119N}
{Nakariakov}, V.~M., \& {Melnikov}, V.~F. 2009, \ssr, 149, 119,
  \dodoi{10.1007/s11214-009-9536-3}

\bibitem[{{Nakariakov} {et~al.}(2003){Nakariakov}, {Melnikov}, \&
  {Reznikova}}]{2003A&A...412L...7N}
{Nakariakov}, V.~M., {Melnikov}, V.~F., \& {Reznikova}, V.~E. 2003, \aap, 412,
  L7, \dodoi{10.1051/0004-6361:20031660}

\bibitem[{{Nakariakov} {et~al.}(2021){Nakariakov}, {Anfinogentov}, {Antolin},
  {Jain}, {Kolotkov}, {Kupriyanova}, {Li}, {Magyar}, {Nistic{\`o}}, {Pascoe},
  {Srivastava}, {Terradas}, {Vasheghani Farahani}, {Verth}, {Yuan}, \&
  {Zimovets}}]{2021SSRv..217...73N}
{Nakariakov}, V.~M., {Anfinogentov}, S.~A., {Antolin}, P., {et~al.} 2021, \ssr,
  217, 73, \dodoi{10.1007/s11214-021-00847-2}

\bibitem[{{Nistic{\`o}} {et~al.}(2014){Nistic{\`o}}, {Pascoe}, \&
  {Nakariakov}}]{2014A&A...569A..12N}
{Nistic{\`o}}, G., {Pascoe}, D.~J., \& {Nakariakov}, V.~M. 2014, \aap, 569,
  A12, \dodoi{10.1051/0004-6361/201423763}

\bibitem[{{Ofman} \& {Liu}(2018)}]{2018ApJ...860...54O}
{Ofman}, L., \& {Liu}, W. 2018, \apj, 860, 54, \dodoi{10.3847/1538-4357/aac2e8}

\bibitem[{{Ofman} {et~al.}(2011){Ofman}, {Liu}, {Title}, \&
  {Aschwanden}}]{2011ApJ...740L..33O}
{Ofman}, L., {Liu}, W., {Title}, A., \& {Aschwanden}, M. 2011, \apjl, 740, L33,
  \dodoi{10.1088/2041-8205/740/2/L33}

\bibitem[{{Ofman} \& {Wang}(2002)}]{2002ApJ...580L..85O}
{Ofman}, L., \& {Wang}, T. 2002, \apjl, 580, L85, \dodoi{10.1086/345548}

\bibitem[{{Pascoe} {et~al.}(2017){Pascoe}, {Goddard}, \&
  {Nakariakov}}]{2017ApJ...847L..21P}
{Pascoe}, D.~J., {Goddard}, C.~R., \& {Nakariakov}, V.~M. 2017, \apjl, 847,
  L21, \dodoi{10.3847/2041-8213/aa8db8}

\bibitem[{{Pascoe} {et~al.}(2013){Pascoe}, {Nakariakov}, \&
  {Kupriyanova}}]{2013A&A...560A..97P}
{Pascoe}, D.~J., {Nakariakov}, V.~M., \& {Kupriyanova}, E.~G. 2013, \aap, 560,
  A97, \dodoi{10.1051/0004-6361/201322678}

\bibitem[{{Pascoe} {et~al.}(2014){Pascoe}, {Nakariakov}, \&
  {Kupriyanova}}]{2014A&A...568A..20P}
---. 2014, \aap, 568, A20, \dodoi{10.1051/0004-6361/201423931}

\bibitem[{{Porter} {et~al.}(1994){Porter}, {Klimchuk}, \&
  {Sturrock}}]{1994ApJ...435..502P}
{Porter}, L.~J., {Klimchuk}, J.~A., \& {Sturrock}, P.~A. 1994, \apj, 435, 502,
  \dodoi{10.1086/174831}

\bibitem[{{Porth} {et~al.}(2014){Porth}, {Xia}, {Hendrix}, {Moschou}, \&
  {Keppens}}]{2014ApJS..214....4P}
{Porth}, O., {Xia}, C., {Hendrix}, T., {Moschou}, S.~P., \& {Keppens}, R. 2014,
  \apjs, 214, 4, \dodoi{10.1088/0067-0049/214/1/4}

\bibitem[{{Roberts} {et~al.}(1983){Roberts}, {Edwin}, \&
  {Benz}}]{1983Natur.305..688R}
{Roberts}, B., {Edwin}, P.~M., \& {Benz}, A.~O. 1983, \nat, 305, 688,
  \dodoi{10.1038/305688a0}

\bibitem[{{Roberts} {et~al.}(1984){Roberts}, {Edwin}, \&
  {Benz}}]{1984ApJ...279..857R}
---. 1984, \apj, 279, 857, \dodoi{10.1086/161956}

\bibitem[{{Shen} {et~al.}(2019){Shen}, {Chen}, {Liu}, {Shibata}, {Tang}, \&
  {Liu}}]{2019ApJ...873...22S}
{Shen}, Y., {Chen}, P.~F., {Liu}, Y.~D., {et~al.} 2019, \apj, 873, 22,
  \dodoi{10.3847/1538-4357/ab01dd}

\bibitem[{{Shen} {et~al.}(2022){Shen}, {Zhou}, {Duan}, {Tang}, {Zhou}, \&
  {Tan}}]{2022SoPh..297...20S}
{Shen}, Y., {Zhou}, X., {Duan}, Y., {et~al.} 2022, \solphys, 297, 20,
  \dodoi{10.1007/s11207-022-01953-2}

\bibitem[{{Shen} {et~al.}(2013){Shen}, {Liu}, {Su}, {Li}, {Zhang}, {Tian},
  {Zhao}, \& {Elmhamdi}}]{2013SoPh..288..585S}
{Shen}, Y.~D., {Liu}, Y., {Su}, J.~T., {et~al.} 2013, \solphys, 288, 585,
  \dodoi{10.1007/s11207-013-0395-4}

\bibitem[{{Shi} {et~al.}(2023){Shi}, {Li}, {Chen}, {Guo}, \&
  {Yuan}}]{2023ApJ...943L..19S}
{Shi}, M., {Li}, B., {Chen}, S.-X., {Guo}, M., \& {Yuan}, S. 2023, \apjl, 943,
  L19, \dodoi{10.3847/2041-8213/acb3c6}

\bibitem[{{Spitzer}(1962)}]{1962pfig.book.....S}
{Spitzer}, L. 1962, {Physics of Fully Ionized Gases}

\bibitem[{{Takasao} \& {Shibata}(2016)}]{2016ApJ...823..150T}
{Takasao}, S., \& {Shibata}, K. 2016, \apj, 823, 150,
  \dodoi{10.3847/0004-637X/823/2/150}

\bibitem[{{Thompson} \& {Myers}(2009)}]{2009ApJS..183..225T}
{Thompson}, B.~J., \& {Myers}, D.~C. 2009, \apjs, 183, 225,
  \dodoi{10.1088/0067-0049/183/2/225}

\bibitem[{{Thompson} {et~al.}(1998){Thompson}, {Plunkett}, {Gurman}, {Newmark},
  {St. Cyr}, \& {Michels}}]{1998GeoRL..25.2465T}
{Thompson}, B.~J., {Plunkett}, S.~P., {Gurman}, J.~B., {et~al.} 1998, \grl, 25,
  2465, \dodoi{10.1029/98GL50429}

\bibitem[{{Van Doorsselaere} {et~al.}(2020){Van Doorsselaere}, {Srivastava},
  {Antolin}, {Magyar}, {Vasheghani Farahani}, {Tian}, {Kolotkov}, {Ofman},
  {Guo}, {Arregui}, {De Moortel}, \& {Pascoe}}]{2020SSRv..216..140V}
{Van Doorsselaere}, T., {Srivastava}, A.~K., {Antolin}, P., {et~al.} 2020,
  \ssr, 216, 140, \dodoi{10.1007/s11214-020-00770-y}

\bibitem[{{Wang} {et~al.}(2021{\natexlab{a}}){Wang}, {Chen}, \&
  {Ding}}]{2021ApJ...911L...8W}
{Wang}, C., {Chen}, F., \& {Ding}, M. 2021{\natexlab{a}}, \apjl, 911, L8,
  \dodoi{10.3847/2041-8213/abefe6}

\bibitem[{{Wang} {et~al.}(2015){Wang}, {Liu}, {Gong}, {Wu}, \&
  {Lin}}]{2015ApJ...805..114W}
{Wang}, H., {Liu}, S., {Gong}, J., {Wu}, N., \& {Lin}, J. 2015, \apj, 805, 114,
  \dodoi{10.1088/0004-637X/805/2/114}

\bibitem[{{Wang} {et~al.}(2009){Wang}, {Shen}, \& {Lin}}]{2009ApJ...700.1716W}
{Wang}, H., {Shen}, C., \& {Lin}, J. 2009, \apj, 700, 1716,
  \dodoi{10.1088/0004-637X/700/2/1716}

\bibitem[{{Wang} {et~al.}(2021{\natexlab{b}}){Wang}, {Ofman}, {Yuan}, {Reale},
  {Kolotkov}, \& {Srivastava}}]{2021SSRv..217...34W}
{Wang}, T., {Ofman}, L., {Yuan}, D., {et~al.} 2021{\natexlab{b}}, \ssr, 217,
  34, \dodoi{10.1007/s11214-021-00811-0}

\bibitem[{{Williams} {et~al.}(2002){Williams}, {Mathioudakis}, {Gallagher},
  {Phillips}, {McAteer}, {Keenan}, {Rudawy}, \&
  {Katsiyannis}}]{2002MNRAS.336..747W}
{Williams}, D.~R., {Mathioudakis}, M., {Gallagher}, P.~T., {et~al.} 2002,
  \mnras, 336, 747, \dodoi{10.1046/j.1365-8711.2002.05764.x}

\bibitem[{{Williams} {et~al.}(2001){Williams}, {Phillips}, {Rudawy},
  {Mathioudakis}, {Gallagher}, {O'Shea}, {Keenan}, {Read}, \&
  {Rompolt}}]{2001MNRAS.326..428W}
{Williams}, D.~R., {Phillips}, K.~J.~H., {Rudawy}, P., {et~al.} 2001, \mnras,
  326, 428, \dodoi{10.1046/j.1365-8711.2001.04491.x}

\bibitem[{{Xia} {et~al.}(2018){Xia}, {Teunissen}, {El Mellah}, {Chan{\'e}}, \&
  {Keppens}}]{2018ApJS..234...30X}
{Xia}, C., {Teunissen}, J., {El Mellah}, I., {Chan{\'e}}, E., \& {Keppens}, R.
  2018, \apjs, 234, 30, \dodoi{10.3847/1538-4365/aaa6c8}

\bibitem[{{Xie} {et~al.}(2019){Xie}, {Mei}, {Huang}, {Lv}, {Roussev}, \&
  {Lin}}]{2019MNRAS.490.2918X}
{Xie}, X., {Mei}, Z., {Huang}, M., {et~al.} 2019, \mnras, 490, 2918,
  \dodoi{10.1093/mnras/stz2576}

\bibitem[{{Yang} {et~al.}(2015){Yang}, {Zhang}, {He}, {Peter}, {Tu}, {Wang},
  {Zhang}, \& {Feng}}]{2015ApJ...800..111Y}
{Yang}, L., {Zhang}, L., {He}, J., {et~al.} 2015, \apj, 800, 111,
  \dodoi{10.1088/0004-637X/800/2/111}

\bibitem[{{Ye} {et~al.}(2021){Ye}, {Cai}, {Shen}, {Raymond}, {Mei}, {Li}, \&
  {Lin}}]{2021ApJ...909...45Y}
{Ye}, J., {Cai}, Q., {Shen}, C., {et~al.} 2021, \apj, 909, 45,
  \dodoi{10.3847/1538-4357/abdeb5}

\bibitem[{{Yuan} {et~al.}(2023){Yuan}, {Fu}, {Cao}, {Ku{\'z}ma}, {Geeraerts},
  {Trelles Arjona}, {Murawski}, {Van Doorsselaere}, {Srivastava}, {Miao},
  {Feng}, {Feng}, {Noda}, {Cobo}, \& {Su}}]{2023NatAs...7..856Y}
{Yuan}, D., {Fu}, L., {Cao}, W., {et~al.} 2023, Nature Astronomy, 7, 856,
  \dodoi{10.1038/s41550-023-01973-3}

\bibitem[{{Zhou} {et~al.}(2020){Zhou}, {Gao}, {Wang}, {Lin}, {Su}, {Jin}, \&
  {Zhang}}]{2020ApJ...905..150Z}
{Zhou}, G., {Gao}, G., {Wang}, J., {et~al.} 2020, \apj, 905, 150,
  \dodoi{10.3847/1538-4357/abc5b2}

\bibitem[{{Zhou} {et~al.}(2021){Zhou}, {Shen}, {Su}, {Tang}, {Zhou}, {Duan}, \&
  {Tan}}]{2021arXiv210902847Z}
{Zhou}, X., {Shen}, Y., {Su}, J., {et~al.} 2021, arXiv e-prints,
  arXiv:2109.02847.
\newblock \doarXiv{2109.02847}

\end{thebibliography}
\bibliographystyle{aasjournal}

\end{CJK*}
\end{document}